\documentclass[conference]{IEEEtran}
\IEEEoverridecommandlockouts

\usepackage{amsmath,amsfonts}

\usepackage{amssymb}
\usepackage{algorithm}
\usepackage{algpseudocode}
\usepackage{graphicx}
\usepackage{bm}
\usepackage{multirow}
\usepackage{subfig}
\usepackage{textcomp}
\usepackage{url}
\usepackage{xcolor}
\usepackage{tikz}
\usepackage{flushend}
\usetikzlibrary{tikzmark}
\usepackage{multirow}
\usepackage{booktabs}

\usepackage{amsthm}

\def\BibTeX{{\rm B\kern-.05em{\sc i\kern-.025em b}\kern-.08em
    T\kern-.1667em\lower.7ex\hbox{E}\kern-.125emX}}
\begin{document}

\title{GPU-Accelerated Vecchia Approximations  of Gaussian Processes for Geospatial Data using Batched Matrix Computations}

\author{
\IEEEauthorblockN{Qilong Pan$^1$, Sameh Abdulah$^2$, Marc G. Genton$^{12}$, David E. Keyes$^2$, Hatem Ltaief$^2$, Ying Sun$^{12}$}
\IEEEauthorblockA{
Division of Computer, Electrical, and Mathematical Sciences and Engineering (CEMSE),\\
$^1$ Statistics Program, \\$^2$ Extreme Computing Research Center, 
\\King Abdullah University of Science and
Technology, \\Thuwal, Jeddah 23955, Saudi Arabia
}
}
\maketitle

\begin{abstract}


Gaussian processes (GPs) are commonly used for geospatial analysis, but they suffer from high computational complexity when dealing with massive data. For instance, the log-likelihood function required in estimating the statistical model parameters for geospatial data is a computationally intensive procedure that involves computing the inverse of a covariance matrix with size $n \times n$, where $n$ represents the number of geographical locations in the simplest case. As a result, in the literature, studies have shifted towards approximation methods to handle larger values of $n$ effectively while maintaining high accuracy. These methods encompass a range of techniques, including low-rank and sparse approximations. Among these techniques, Vecchia approximation is one of the most promising methods to speed up evaluating the log-likelihood function. This study presents a parallel implementation of the Vecchia approximation technique, utilizing batched matrix computations on contemporary GPUs. The proposed implementation relies on batched linear algebra routines to efficiently execute individual conditional distributions in the Vecchia algorithm. We rely on the KBLAS linear algebra library to perform batched linear algebra operations, reducing the time to solution compared to the state-of-the-art parallel implementation of the likelihood estimation operation in the {\it ExaGeoStat} software by up to \textcolor{black}{700X, 833X, 1380X} on 32GB GV100, 80GB A100, and 80GB H100 GPUs, respectively, with the largest matrix dimension that can fully fit into the GPU memory in the dense Maximum Likelihood Estimation (MLE) case. We also successfully manage larger problem sizes on a single NVIDIA GPU, accommodating up to 1 million locations with 80GB A100 and H100 GPUs while maintaining the necessary application accuracy. We further assess the accuracy performance of the implemented algorithm, identifying the optimal settings for the Vecchia approximation algorithm to preserve accuracy on two real geospatial datasets: soil moisture data in the Mississippi Basin area and wind speed data in the Middle East.

\end{abstract}


\section{Introduction}
Gaussian Processes (GPs) play a crucial role in spatial statistics applications, where they are employed for modeling and predicting geospatial data. This is achieved by defining the mean and covariance functions of the process within a region. For GPs, a parametric form of the covariance function defines the correlation between the spatial locations using a set of parameters, thereby characterizing the dependence of the spatial data. GPs are employed in many applications, such as geostatistics, machine learning, and computer vision. However, a significant challenge arises when dealing with large datasets collected at irregularly spaced locations where the computational complexity of the GP modeling and prediction increases cubically with the number of spatial locations. This poses a limitation for applications with large spatial datasets.

Numerous studies have addressed the computational issues associated with large-scale GP modeling and prediction. Most of the efforts have centered on two main directions: sparse approximation and low-rank approximation to the covariance matrix. Examples of the former include covariance tapering methods ~\cite{furrer2006covariance,kaufman2008covariance,bevilacqua2016covariance}. In covariance tapering, the covariance function is multiplied by a tapering function that decays to zero as the distance between two locations increases. This process produces from the original dense covariance method a sparse one that can be managed less cumbersomely. Other studies assume the sparsity of the original covariance matrix by partially including correlation between some spatial locations and ignoring others, i.e., sparse inverse covariance methods~\cite{nychka2015multiresolution,abdulah2018exageostat}. Moreover, the emergence of modern hardware architectures that support low-precision computation, such as NVIDIA GPUs, has facilitated the optimization of sparse inverse covariance methods by applying different precisions to various parts of the dense covariance matrix to reduce the computational complexity instead of ignoring them~\cite{abdulah2019geostatistical,abdulah2021accelerating,cao2022reshaping}. For the latter, different types of low-rank approximations are exploited, which allows faster computation and less memory consumption compared to the original dense matrix~\cite{katzfuss2011spatio,huang2018hierarchical,abdulah2018parallel, mondal2022parallel}.

Vecchia approximation is one of the earliest GP statistical approximation methods. It involves replacing the high-dimensional joint distribution of the GP with a product of univariate conditional distributions. A small set of observations is conditioned in each conditional distribution, as described in~\cite{vecchia1988estimation}. This method involves a portion of the locations at once instead of all the locations in the modeling process, allowing faster execution and less memory consumption. Vecchia approximation results in an approximated log-likelihood function with a computational complexity of $\mathcal{O}(nm^3)$ instead of the standard $\mathcal{O}(n^3)$ complexity. Here, $n$ denotes the number of spatial locations, and $m$ represents the number of neighbors considered in the conditional distributions, which is significantly smaller than $n$. Another advantage of the Vecchia approximation is that it is amenable to parallel computing since terms may be computed independently. However, a challenge to scaling Vecchia approximation is that it requires small matrix operations, which can be more suitable to parallelize on CPUs rather than GPUs. While GPUs are intended to solve problems with huge computational requirements, numerous applications, including Vecchia approximation, offer many small tasks instead. For reviews in Vecchia approximation, see~\cite{katzfuss2021general, katzfuss2022scaled, zhang2022multi, jimenez2023scalable}.

The latest TOP500 Supercomputers list released in November 2023 reveals that 9 of the top 10 supercomputers worldwide use NVIDIA, Intel, or AMD GPU accelerators, allowing peak performance levels of more than 1.6 ExaFlops/s~\cite{top500}. GPUs are favored to accelerate tasks because of their superior computational power and energy efficiency compared to CPUs. With CPUs in complex matrix operations, accelerators have traditionally been employed to handle the computation. GPUs are typically leveraged for computationally-intensive tasks, while CPUs are better suited for latency-sensitive ones. However, this approach does not work well with small matrix operations that do not fully utilize the existing accelerators. Instead, concurrent batched operations can be used to execute the same operation across multiple small matrices on a single GPU to allow better exploitation of the existing hardware~\cite{haidar2015batched,abdelfattah2019fast}.

In this work, we leverage the power of modern GPU architectures to accelerate the Vecchia approximation algorithm of the Gaussian field using batched matrix operations. The approach applies uniform operations to batches of small matrices to leverage the underlying GPU accelerators. We assess the execution and accuracy performance of our implementation on three different NVIDIA GPU accelerators, GV100, A100, and H100, showing fast execution using batched matrix operations with an accuracy comparable to the dense solution provided with the state-of-the-art Gaussian process software, i.e., {\it ExaGeoStat}~\cite{abdulah2018exageostat}. This study also shows we can handle larger problem sizes with Vecchia approximation on a single GPU compared to using exact likelihood. In our numerical study and real dataset analysis, we find that the suitable number of neighbors (conditioning set) required for modeling each location is at most $60$. Vecchia approximation effectively reduces the memory complexity of the MLE operation from $\mathcal{O}(n^2)$ to $\mathcal{O}(nm^2)$, e.g., $m\leq60$. In the experimental section, we assess the performance and accuracy of our implemented approach, emphasizing the benefits of utilizing Vecchia approximation over the exact likelihood while maintaining the necessary accuracy.

The paper is structured as follows: In Section \ref{sec:contribution}, we summarize our contributions. In Section~\ref{sec:related}, we review related work. Section~\ref{sec:background} provides a comprehensive background for the paper. Section~\ref{sec:framework} offers a detailed explanation of our proposed implementation. Section~\ref{sec:exp} presents the evaluation of our implementation from both accuracy and performance perspectives, and we conclude in Section~\ref{sec:conclusion}.

\section{Contributions}
\label{sec:contribution}
We summarize the contributions of the paper as follows:

\begin{itemize}
    \item We introduce a GPU-accelerated implementation of the well-known Vecchia approximation algorithm for estimating statistical model parameters in the context of climate and weather applications.
    \item We utilize the KBLAS library and batched linear algebra operations to enhance the speed of our implementation on contemporary GPU architectures, including those from NVIDIA, such as GV100, A100, and H100.
    \item We assess the accuracy of the proposed implementation through numerical study and two real datasets: a soil moisture dataset from the Mississippi Basin area and a wind speed dataset from the Middle East region. We emphasize identifying the optimal settings that allow the Vecchia algorithm to achieve performance on par with the exact MLE operation as implemented in state-of-the-art HPC geostatistics software, i.e., {\it ExaGeoStat}.
    \item We assess the execution performance of the GPU-based Vecchia algorithm on three different NVIDIA GPU architectures, achieving speedups of up to \textcolor{black}{700X, 833X, 1380X} on 32GB GV100, 80GB A100, and 80GB H100 GPUs, respectively, compared to the exact MLE operation.
    \item Our implementation accommodates larger problem sizes within the same GPU memory, enabling improved modeling for high-resolution geospatial data.
\end{itemize}

\section{Related Work}
\label{sec:related}
\subsection{Vecchia Approximation}
The Vecchia method, as described in the study on Gaussian process estimation ~\cite{vecchia1988estimation}, has been investigated and proven to be computationally feasible for non-gridded spatial data. The basic idea behind the Vecchia approximation is to approximate the full covariance matrix of the Gaussian process by considering a smaller subset of the data points and using a conditional independence assumption. This approximation is particularly useful in cases where the full covariance matrix is too large to handle efficiently. The study \cite{guinness2018permutation} offers an in-depth examination of the Vecchia method, highlighting its effectiveness in handling spatial data with various characteristics. In \cite{guinness2018permutation}, different spatial orderings are investigated to enhance the approximation of Gaussian processes via the Vecchia algorithm. Specifically, maximum–minimum distance and random orderings demonstrate a remarkable 99\% relative efficiency of the approximation algorithm while requiring only a minimal set of 30 neighboring data points. 
Furthermore, an idea of grouping locations that share the most common neighbors has been introduced to expedite computations while gaining accuracy. Subsequently, to adapt the Vecchia approximation for MLE operation, the Fisher scoring optimization algorithm \cite{guinness2021gaussian} has been employed, resulting in MLE convergence within a few iterations. A dedicated R package, GpGp \cite{guinness2021gaussian}, has also been developed to facilitate these computations. The GpGp team achieved victory in recent competitions assessing the effectiveness of existing software in statistical parameter estimation and prediction, thanks to their utilization of the GpGp package~\cite{huang2021competition, abdulah2022second,hong2023third}.

A general framework for the Vecchia approximation has been introduced \cite{katzfuss2021general}. This framework unifies the estimation, prediction, and emulation with the Vecchia approximation while establishing seamless integration with other Gaussian approximation methodologies. The method is primarily deployed to approximate the likelihood of statistical models, particularly in scenarios characterized by expensive computations and complicated modelings. Notable applications include addressing intractable spatial extremes models \cite{huser2022vecchia}, optimizing Bayesian processes at a large scale \cite{jimenez2023scalable}, and advancing the compositional warpings to construct nonstationary spatio-temporal covariance models \cite{vu2023constructing}. These applications underscore the versatility of the Vecchia method in handling a range of challenging statistical problems.


\subsection{GPU-based Acceleration in Spatial Statistics}
GPU accelerators have been employed in spatial statistics to tackle various problems, aiming to efficiently process large-scale data and maximize data utilization for parameter estimation in statistical models. For instance, \cite{zhang2015large} primarily concentrates on data parallel approaches for tasks such as spatial indexing, spatial joins, polygon rasterization, decomposition, and point interpolation. These approaches are further expanded to encompass distributed computing nodes by integrating multiple GPU implementations. 

Another example is \cite{li2014accelerating}, which focuses on accelerating the computation of high-order spatial statistics, particularly in geology, by introducing a GPU-based parallel approach, significantly enhancing the computational efficiency of high-order spatial statistics. The key feature of this approach is the utilization of template-stage parallelism. In a real-world application involving a large-scale dataset, \cite{zhang2015efficient} tackles the significant computational challenge of handling over 375 million species occurrence records sourced from the global biodiversity information facility by leveraging GPUs. Their research is notable for the impressive performance enhancements attained by applying parallel zonal statistics. Furthermore, \cite{zhang2017gpu} presents a novel GPU-accelerated approach for adaptive kernel density estimation to address the computational challenges related to bandwidth determination in spatial point pattern analysis. This method incorporates optimizations that reduce algorithmic complexity and harness GPU parallelism, resulting in significant speed improvements. \cite{prasad2015vision} also underscores the importance of expediting geospatial computations and analytics through shared and distributed memory parallel platforms. They advocate using GPUs, which boast hundreds to thousands of processing cores for parallelization. When working with spatial Gaussian data, these datasets are typically represented as realizations derived from a Gaussian spatial random field. The GPU-based parallel solutions presented in their work underscore the promising potential of GPU technology in this field.

\section{Background}
\label{sec:background}

\subsection {Gaussian Process and Likelihood Function}
Statistical modeling involves analyzing correlations among various spatial or spatio-temporal points distributed regularly or irregularly over a geographical area. In the purely spatial scenario, these datasets are assumed to be realizations observed from a Gaussian spatial random field. Consider a set of $n$ locations, ${\bf s}_1, \ldots, {\bf s}_n \in \mathbb{R}^d$, and their observations, $\mathbf{y} = \left(y_1, \ldots, y_n\right)^\top$ which is indicated by $y_i:=y\left({\bf s}_i\right) \in \mathbb{R}$. We model the data $\mathbf{y}$ with Gaussian process, $\mathbf{y} \sim \mathcal{N}\left(\bm \mu, \bm \Sigma_{\boldsymbol\theta}\right)$, where $\bm \Sigma_{\boldsymbol\theta}$ is a covariance matrix with $(i, j)$ entry determined by a given covariance function $C_{\boldsymbol\theta}\left({\bf s}_i, {\bf s}_j\right)$ relying on a vector of covariance parameters,  ${\boldsymbol\theta}$.  Without loss of generality, we assume that $\mathbf{y}$ has a mean of zero. Statistical inference about $\boldsymbol{\theta}$ is often based on the Gaussian log-likelihood function, which entails a cubic computational complexity~\cite{abdulah2018exageostat}:
\begin{equation}
	\label{eq:likeli}
	\ell({\boldsymbol\theta};\mathbf{y})=-\frac{n}{2}\log(2\pi) - \frac{1}{2}\log |{{\boldsymbol \Sigma}({\boldsymbol\theta})}|-\frac{1}{2}{\mathbf{y}}^\top {\boldsymbol \Sigma}({\boldsymbol\theta})^{-1}{\mathbf{y}}.
\end{equation}

The maximum likelihood estimator of $\boldsymbol\theta$ is the value $\hat{\boldsymbol\theta}$ that maximizes $\ell({\boldsymbol\theta})$ in the equation~(\ref{eq:likeli}). Examples of existing covariance functions are the isotropic Matérn covariance function~\cite{wang2023parameterization} in (\ref{eq:maternkernel}) and the power exponential kernel in (\ref{eq:powexpkernel}):
\begin{align}
    C({\bf s}_i, {\bf s}_j)&=\sigma^2 \frac{2^{1-\nu}}{\Gamma(\nu)}\left(\frac{\|{\bf s}_i-{\bf s}_j\|}{\beta}\right)^\nu K_\nu\left(\frac{\|{\bf s}_i-{\bf s}_j\|}{\beta}\right),
    \label{eq:maternkernel}\\
    C({\bf s}_i, {\bf s}_j)&=\sigma^2 \exp \left(-\frac{\|{\bf s}_i-{\bf s}_j\|^\nu}{\beta}\right),
    \label{eq:powexpkernel}
\end{align}
where ${\boldsymbol\theta} = \left(\sigma^2, \beta, \nu \right)^\top$, $\sigma^2$ is the variance, $\Gamma(\cdot)$ is the gamma function, $K_\nu(\cdot)$ is the Bessel function of the second kind of order $\nu$, and $\beta>0$ and $\nu>0$ are range and smoothness parameters, respectively. 

\subsection{Kullback-Leibler (KL) Divergence}
KL divergence is a statistical metric to measure the difference between two given probability distributions. Assuming two models $P$ and $Q$,  the KL divergence of $P$ from $Q$ is the amount of information lost when using $Q$ as a model compared to the actual distribution $P$. It is represented as $D_{\text{KL}}(P\parallel Q)$, with $P$ and $Q$ representing the two probability distributions under comparison. KL divergence is used in various fields, including information theory, machine learning, and statistics.

For two distributions, $P$ and $Q$ of a continuous random variable, KL divergence is defined as, 
$$ D_{\text{KL}}(P\parallel Q)=\int _{-\infty }^{\infty }p(x)\log \left({\frac {p(x)}{q(x)}}\right)\,dx, 
$$
where $p$ and $q$ denote the probability densities of $P$ and $Q$. In the Gaussian fields, $P = \mathcal{N}_0(\boldsymbol 0, \boldsymbol \Sigma_0)$ and $Q = \mathcal{N}_1(\boldsymbol 0, \boldsymbol \Sigma_1)$, the KL divergence of $\mathcal N_0$ and $\mathcal N_1$ is \cite{duchi2007derivations}: 
\begin{equation}
    D_{\text{KL}}({\mathcal {N}}_{0}\parallel {\mathcal {N}}_{1})={\frac{1}{2}}\left\{\operatorname {tr} \left({ {\boldsymbol \Sigma }}_{1}^{-1}{ {\boldsymbol \Sigma }}_{0}\right)-k+\log {|{ {\boldsymbol \Sigma }}_{1}| \over |{{\boldsymbol \Sigma }}_{0}|}\right\},
    \label{eq:kl-gaussian}
\end{equation}
where $\boldsymbol \Sigma_0 := \boldsymbol \Sigma(\boldsymbol\theta_0)$ and $\boldsymbol \Sigma_1 := \boldsymbol \Sigma(\boldsymbol\theta_1)$. 

In this paper, we use the KL divergence criterion to assess the loss of information when relying on the Vecchia approximation algorithm compared to the exact Gaussian likelihood. Thus, with the plug-in of the Vecchia approximated Gaussian likelihood in (\ref{eq:kl-gaussian}), $Q = \mathcal{N}_a(\boldsymbol 0, \boldsymbol\Sigma_a)$ where $\boldsymbol \Sigma_a$ represents an approximated covariance matrix to $\boldsymbol \Sigma_0$, (\ref{eq:kl-gaussian}) is simplified as,
\begin{equation}
    D_{\text{KL}}({\mathcal {N}}_{0}\parallel {\mathcal {N}}_{a})= \ell_0({\boldsymbol\theta};\boldsymbol{0}) - \ell_a({\boldsymbol\theta};\boldsymbol{0}),
    \label{eq:kl-gaussian-vecchia}
\end{equation}
where $\ell_0({\boldsymbol\theta};\boldsymbol{0})$ is the exact log-likelihood at the $\mathbf{y} = \boldsymbol{0}$ while $\ell_a({\boldsymbol\theta};\boldsymbol{0})$ is the Vecchia-approximated log-likelihood at $\mathbf{y} = \boldsymbol{0}$. 

\subsection{Batched Linear Algebra Computations}
Considering the case of block-sparse matrices, non-zero elements are clustered in different parts of the matrix, providing an alternative approach to handling them compared to using dense or sparse linear solvers. The size of these dense blocks may suggest treating them as a collection of small, dense matrices instead. In the literature, performing identical operations on multiple small dense matrices is known as batched execution. Therefore, the concept of batched linear algebra routines is often employed in various applications where the same operations can be applied simultaneously to multiple small, dense matrices. These operations are performed independently on each matrix, enabling more efficient use of computational resources, particularly on high-performance hardware like GPUs.

Numerous software libraries have facilitated linear algebra operations in dense and approximate formats. These include
LAPACK~\cite{anderson1999lapack}, SLATE~\cite{gates2019slate}, PLASMA~\cite{dongarra2019plasma}, HiCMA~\cite{abdulah2019hierarchical}, H2Opus~\cite{zampini2022h2opus}, MAGMA~\cite{agullo2009numerical}, KBLAS~\cite{abdelfattah2016kblas}, and BLIS~\cite{van2015blis}.  Several of these libraries offer methods to enhance efficient batched processing on both CPUs and GPUs, including batched routines for various Basic Linear Algebra Subprograms (BLAS) and LAPACK operations. The Batched BLAS extension enhances the traditional BLAS library by enabling simultaneous processing of multiple small matrices (up to 1024). Its main goal is to diminish the computational burden caused by frequent function calls. Designed for modern hardware, it optimally utilizes the parallel processing capabilities of multi-core processors and GPUs \cite{ dong2016magma,vu2023constructing}.

\begin{table}[h]
\caption{The impact of three data layouts}
\label{tab:my-table}
\centering
\begin{tabular}{lll}
\toprule
           & Pros                     & Cons                                \\\midrule
P2P        & Flexible appendage       & Noncontiguous data          \\
Strided    & Contiguous data  & Expensive appendage                 \\
Interleave & Vectorization            & Complexity format \\ \bottomrule
\end{tabular}
\end{table}

In this context, our focus is directed toward GPU-batched operations. Presently, three commonly used libraries for conducting batched operations are cuBLAS, MAGMA \cite{dong2016magma}, and KBLAS \cite{abdelfattah2016kblas}. It is essential to note that each of the three libraries offers distinct data layouts and functions. The functions are essential to the Vecchia algorithm, and the data layout greatly impacts the performance. The distinction becomes especially significant when considering the choice of data layout, which can be pointer-to-pointer (P2P), strided, or interleaved formats, as shown in Table \ref{tab:my-table}. The P2P data layout offers the flexibility to append additional data at the end of the batched data array by adding a pointer. However, this convenience is offset by the memory loading burden stemming from the non-contiguous nature of the small matrices, which are scattered across the global memory on the GPU. Conversely, the strided data layout allocates contiguous memory for the entire collection of small matrices, thus resulting in efficient access to these matrices. Nevertheless, adding extra data incurs significant expense in terms of memory usage. On the other hand, the interleaved layout is distinguished by its ability to utilize memory through vectorization fully. However, this benefit is primarily realized for very small matrices (less than 24) \cite{dongarra2017design}.






\section{Batched Vecchia Approximation}
\label{sec:framework}

This section offers an in-depth description of our proposed framework for the batched Vecchia algorithm. It commences by explaining the preprocessing step, which entails reordering the location set and selecting the nearest neighbors for each location from previous locations. Subsequently, we delve into the memory requirements for the batched Vecchia algorithm with a comprehensive description of the proposed implementation. Finally, we compare the expected memory usage and computational complexity of the Vecchia algorithm in contrast to the exact MLE solution.

\subsection{Reorderings}

The initial step of the Vecchia algorithm involves efficiently reordering the locations to identify nearest-neighbor points for each location. Consequently, selecting the reordering method plays a crucial role in ensuring the accuracy of the log-likelihood approximation. This is because, for a given observation, the selection of nearest neighbors from prior observations highly depends on the sequence in which the data is arranged. According to \cite{guinness2021gaussian}, the maximum–minimum distance (Maxmin) ordering method is identified as suitable for the Vecchia algorithm. Notably, random ordering achieves comparable accuracy to Maxmin, as demonstrated in \cite{guinness2018permutation}. However, when dealing with large-scale problems, employing the Maxmin ordering method can become impractical due to its computational complexity, which can scale as $\mathcal{O}(n^3)$. Furthermore, the GpGp R package recommends random ordering for scenarios where locations exceed 100K \cite{guinness2021gaussian}. Consequently, random ordering has been used in our experiments. In addition to random ordering, the Morton ordering method is also considered, given its efficiency in tile-low rank Cholesky factorization as reported in \cite{akbudak2017tile}.

\begin{figure}
    \centering
    \subfloat[45th in Random Ordering]{\includegraphics[width=0.23\textwidth]{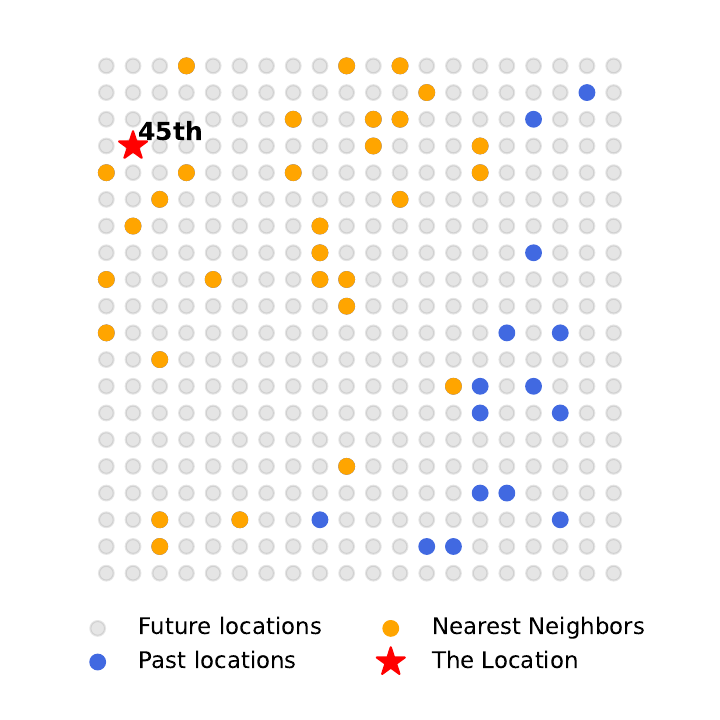}}
    \hspace{4mm}
    \subfloat[250th in Random Ordering]{\includegraphics[width=0.23\textwidth]{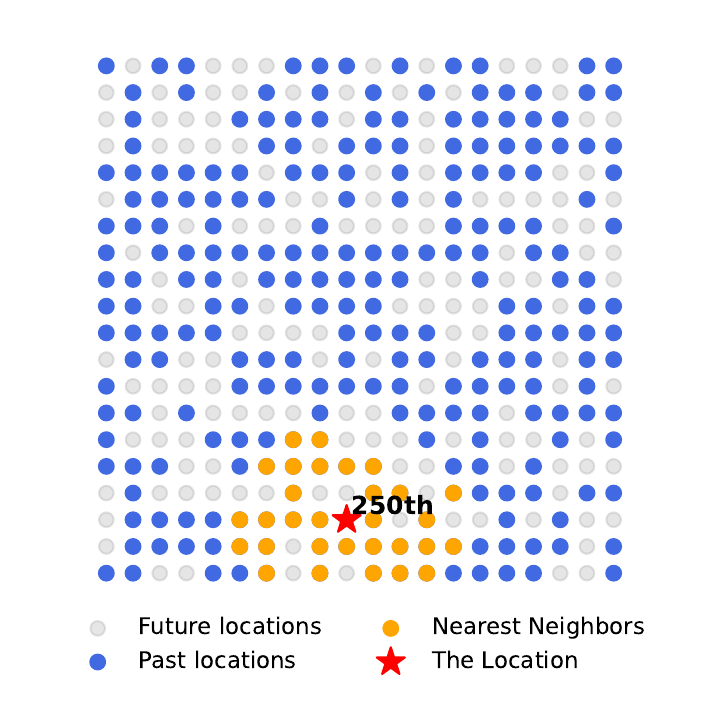}}\\
    \vspace{-5mm}
    \subfloat[45th in Morton Ordering]{\includegraphics[width=0.23\textwidth]{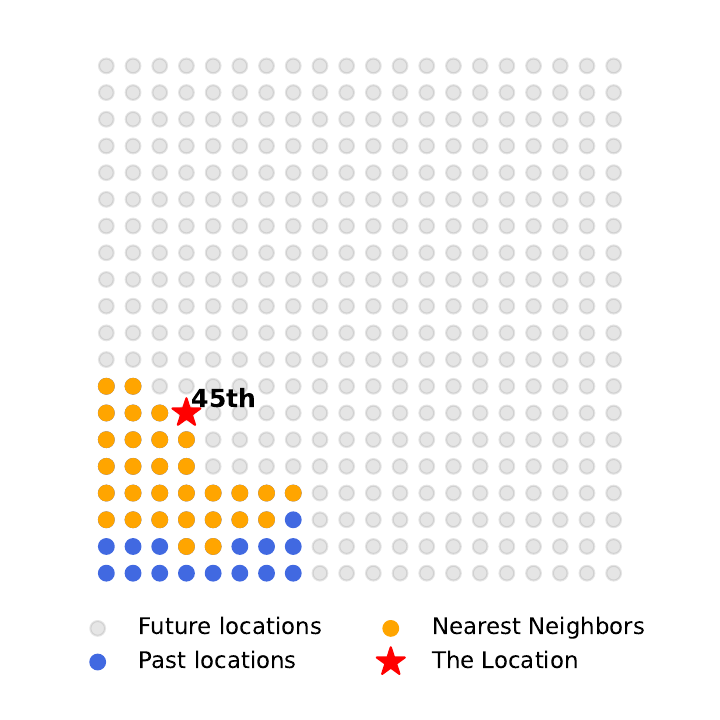}}
    \hspace{4mm}
    \subfloat[250th in Morton Ordering]{\includegraphics[width=0.23\textwidth]{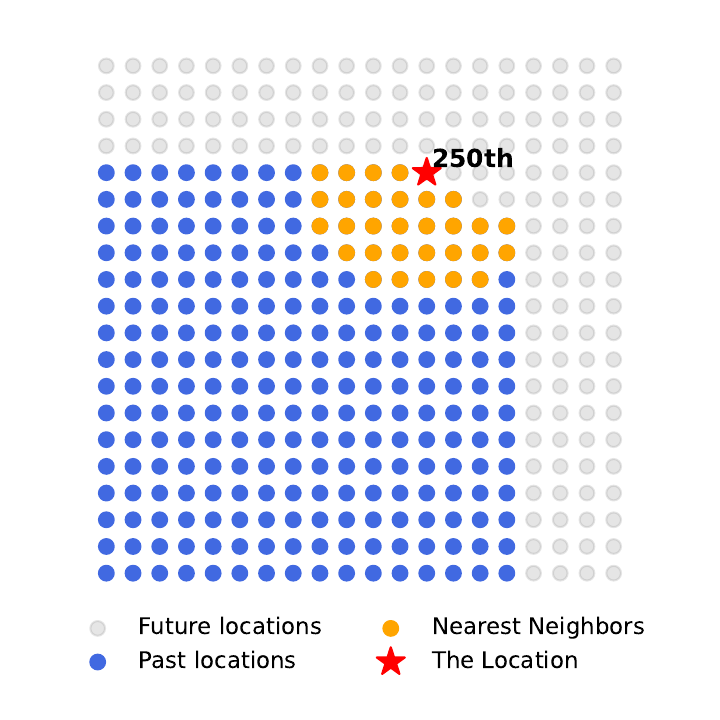}}
    \caption{The example of random and Morton ordering on locations $20 \times 20$. (First row) The 45th and 250th locations (red stars) in the random ordering are marked with their nearest neighbors (orange circle); (Second row) The 45th and 250th locations (red stars) in the Morton order algorithm are marked with their nearest neighbors (orange circle). Blue circles indicate past locations for a given ordering algorithm.}
    \label{fig:ordering}
\end{figure}

To elucidate the impacts of two distinct ordering methods on finding the best nearest neighbors, we present an illustrative example involving a grid of $20 \times 20$ locations, as depicted in Figure \ref{fig:ordering}. This example demonstrates that, in contrast to the Morton ordering method, random ordering confers an advantage for locations sequenced toward the end of the ordering process. It retains the nearest neighbors immediately surrounding a target location. However, this comes initially at the expense of losing proximity accuracy for locations ordered. In the experimental section, we will apply both methods to reorder our locations and empirically demonstrate which can yield superior results.

\subsection{Batched Vecchia Approximated Likelihood Algorithm}

In an analysis involving a set of geospatial data points $\mathbf{y}$ across various locations, and their observed values, when we apply any reordering $\tau$ to the sequence of these locations, the likelihood expression for the dataset retains its form when represented as a sequence of conditional densities, 
\begin{align}
    L({\boldsymbol\theta};\mathbf{y}) &= p_{\boldsymbol\theta}\left(y_1, \ldots, y_n\right) \\
    & =p_{\boldsymbol\theta}\left(y_1^\tau\right) \prod_{i=2}^n p_{\boldsymbol\theta}\left(y_i^\tau \mid y_1^\tau, \ldots, y_{i-1}^\tau\right).
\end{align}
Vecchia approximation is expressed as (\ref{eq:vecchia-vector}) and it replaces the conditioning vectors $\left(y_1^\tau, \ldots, y_{i-1}^\tau\right)$ with  $(y_{j_{i 1}}^\tau, \ldots, y_{j_{im_i}}^\tau)$ which are nearest neighbors to the target $y_{j_{i}}^\tau$; $J_i$ is defined as $\left\{j_{i 1}, \ldots, j_{i m_i}
\right\}$ and $J=\left\{J_1, \ldots, J_n\right\}$. \cite{guinness2018permutation}
\begin{align}
    p_{\boldsymbol\theta, \tau, J}\left(y_1, \ldots, y_n\right)
    &=
    p_{\boldsymbol\theta}\left(y_1^\tau\right) \prod_{i=2}^n p_{\boldsymbol\theta}\left(y_i^\tau \mid y_{j_{i 1}}^\tau, \ldots, y_{j_{im_i}}^\tau\right) \\
    &= 
    p_{\boldsymbol\theta}\left(y_1^\tau\right) \prod_{i=2}^n p_{\boldsymbol\theta}\left(y_i^\tau \mid \mathbf{y}^{\tau}_{J_i} \right).\label{eq:vecchia-vector}
\end{align}
The accuracy of the Vecchia algorithm's approximation is dependent on the choice of permutation $\tau$ and $J$, because the permutation $\tau$ creates specific ordering from 2D to 1D and the $J$ represents the deviations from the exact likelihood. \cite{guinness2018permutation} \cite{vecchia1988estimation}

\begin{figure}[h]
    \centering
    \includegraphics[width=0.48\textwidth]{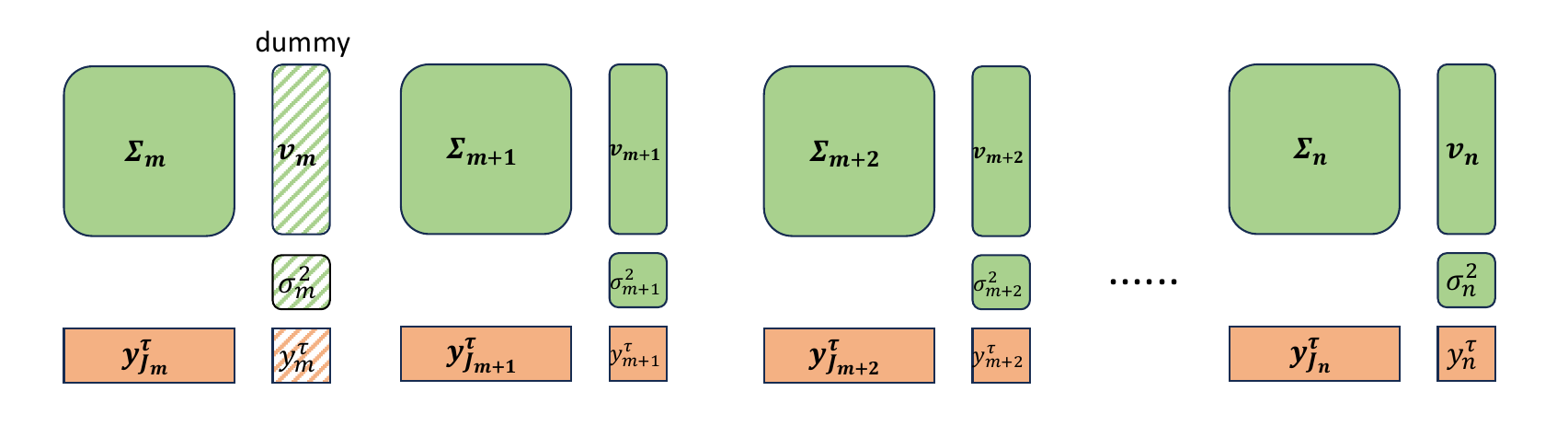}
    \caption{Batched Vecchia algorithm description. $\bm \Sigma_{m:n}$ are constructed by the nearest neighbors of $\mathbf y^{\tau}_{m:n}$. The batched POTRF routine is applied to these matrices. After this decomposition, the resulting outputs are utilized as inputs for the batched TRSV operation with $\mathbf v_{m:n}$ and $\mathbf y^{\tau}_{\mathbf J_{m:n}}$, separately. }
    \label{fig:alg-descrip}
\end{figure}

\begin{figure}[h]
    \centering
    \includegraphics[width=0.3\textwidth]{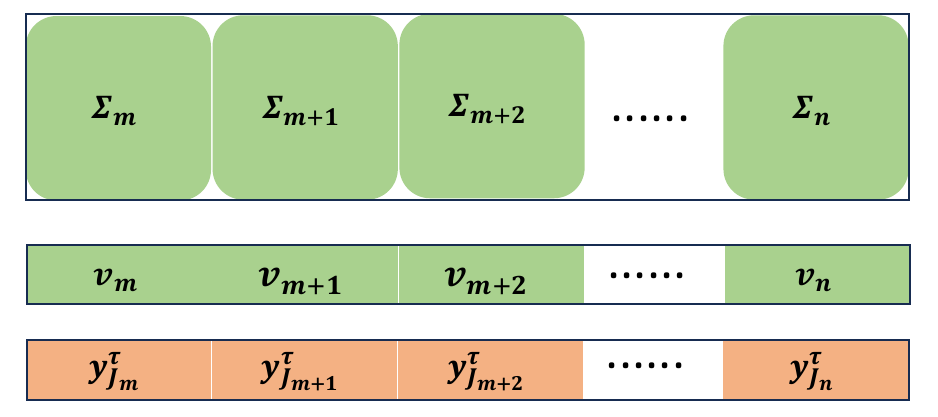}
    \caption{Contiguous data allocation in the GPU global memory.}
    \label{fig:data-structure}
\end{figure}

To implement the Vecchia algorithm, for each spatial location, we must compute a covariance matrix of its nearest neighbors and a cross-covariance vector between the location and its nearest neighbors. Figure \ref{fig:alg-descrip} depicts the required scalar/vector/matrix for each spatial location, while Figure \ref{fig:data-structure} shows the contiguous data allocation utilized in the hardware. $y^{\tau}_{i}$ and $\mathbf{y}^{\tau}_{J_i}$  (orange) are the $i$th observation and its neighbors' observations which exactly match the representation in equation (\ref{eq:vecchia-vector}); $\sigma_i$ and  $\bm \Sigma_i$ (green) are the corresponding variance and covariance matrix for the $i$th observation and its neighbors; $\mathbf{v}_i$ (green) is the cross-covariance matrix of $y^{\tau}_{i}$ and $\mathbf{y}^{\tau}_{J_i}$. For every pair $(y^{\tau}_{i}, \mathbf{y}^{\tau}_{J_i}, \sigma_i, \bm \Sigma_i)$, their log-likelihoods, $p_{\boldsymbol\theta}\left(y_i^\tau \mid \mathbf{y}^{\tau}_{J_i} \right)$, are independent of each other. That is to say, the task of computing log-likelihood in the (\ref{eq:vecchia-vector}) can be divided into $n$ independent tasks. In the log-likelihood computation, the batched operations are conducted with regard to $\mathbf{v}_i$,  $\bm \Sigma_i$ (green), $\mathbf{y}^{\tau}_{J_i}$  (orange) where the two dummy dashed vectors are not used for log-likelihood calculation but purely for simplifying the batched algorithm. 

\begin{algorithm}[t]
\caption{Batched Vecchia algorithm}\label{alg:batchedvecchia}
\begin{algorithmic}[1]
\State \textbf{Input:} $m$, $n$, $\tau$, $C(\cdot, \cdot)$
\State \textbf{Output:} $\ell$ (log-likelihood)
\State $\tau:\{(\mathbf{s}_1, y_1),  \ldots, (\mathbf{s}_n, y_n)\} \rightarrow$ $\{(\mathbf{s}_1^\tau, y_1^\tau),  \ldots, (\mathbf{s}_n^\tau, y_n^\tau)\}$ \Comment{Permutation}
\While{$m+1\leq j \leq n$} \Comment{Nearest neighbors}
\State $J_{j} \gets mNearstNeighbors(\mathbf{s}_1^{\tau}, \mathbf{s}_2^{\tau}, \ldots, \mathbf{s}_{j-1}^{\tau};\mathbf{s}_{j}^{\tau}, m)$
\EndWhile
\State 
\textcolor{blue}{
\State $\bm \sigma_{m:n} \gets batchC(\mathbf{s}_{m:n}^{\tau}, \mathbf{s}_{m:n}^{\tau})$ \Comment{Batched Kernel}
\State $\bm \Sigma_{m:n} \gets batchC(\mathbf{s}_{J_{m:n}}^{\tau}, \mathbf{s}_{J_{m:n}}^{\tau})$
\State $\mathbf{v}_{m:n} \gets batchC(\mathbf{s}_{m:n}^{\tau}, \mathbf{s}_{J_{m:n}}^{\tau})$
\State $\mathbf{y}^{\tau}_{J_{m:n}}$
}
\State $\bm\sigma^{old} \gets \left(\sigma_{m}, \ldots, \sigma_{n}\right)^T$
\textcolor{blue}{
\State $ \bm L_{m:n} \gets batchPOTRF(\bm \Sigma_{m:n})$ \Comment{Batched operations}
\State $ \mathbf{v}'_{m:n} \gets batchTRSV(\bm L_{m:n}, \mathbf{v}_{m:n})$
\State $ \mathbf{y'}^{\tau}_{j_{m:n}} \gets batchTRSV(\bm L_{m:n}, \mathbf{y}^{\tau}_{j_{m:n}})$
}
\State $\bm Y' = (\mathbf{y}^{\tau}_{j_m}, \mathbf{y}^{\tau}_{j_{m+1}}, \ldots, \mathbf{y}^{\tau}_{j_n})$ \Comment{Concatenate}
\State $\bm V' = (\mathbf{v}_m, \mathbf{v}_{m+1}, \ldots, \mathbf{v}_n)$ 
\State $\mathbf{v}_m \gets \mathbf{y}^{\tau}_{j_m}$
\textcolor{blue}{
\State $\bm \mu' \gets DotProduct(\bm Y'^T, \bm V')$ \Comment{Correction vectors}
\State $\bm \sigma' \gets DotProduct(\bm V'^T,
\bm V')$
}
\State $\ell_m \gets -computeLogDet(\bm L_m) - \frac{\mu'_m}{2} - \frac{m}{2}log(2\pi)$
\State $ \bm\mu^{new} \gets \bm\mu'_{(m+1):n}$\Comment{Vecchia updates}
\State $ \bm\sigma^{new} \gets \bm\sigma^{old} - \bm\sigma'_{(m+1):n}$
\While{$ (m+1)\leq i \leq n$} \Comment{Univariate Gaussian $\ell$}
\State $\ell_i \gets -\frac{1}{2} \left( \left(  \frac{x-\mu^{new}_{i}}{\sigma^{new}_{i}}\right)^2 + log(2\pi) + 2log(\sigma^{new}_{i}) \right)$
\EndWhile
\State $\ell \gets \ell_m + \ell_{(m+1)} + \ldots + \ell_n$
\end{algorithmic}
\end{algorithm}

Algorithm~\ref{alg:batchedvecchia} provides a high-level overview of the batched operations employed in the Vecchia approximation, as depicted in Figure~\ref{fig:alg-descrip}. The inputs are $m$, representing the size of the conditioning set, $n$, the total number of observations, $\tau$, the permutation set of the locations, and $C(\cdot, \cdot)$, the specified covariance function. The output is the log-likelihood estimation $\ell$. The following points outline the specific low-level implementation details:

\begin{itemize}
    \item \textit{GPU acceleration:} In Algorithm \ref{alg:batchedvecchia}, lines 9 and 10 describe the utilization of a \textcolor{black}{batched CUDA kernel} to generate the covariance matrix \cite{geng2023gpu}, which \textcolor{black}{overcomes the primary computational bottleneck} in Vecchia approximation as discussed in \cite{guinness2018permutation}. Following this, we leverage batched operations, notably the Cholesky decomposition for symmetric matrices (POTRF) and the subsequent use of a triangular linear solver (TRSV), as demonstrated in lines \textcolor{black}{13, 14, and 15}, using the KBLAS library \cite{abdelfattah2016kblas}. In lines \textcolor{black}{19 and 20}, we introduce a CUDA kernel implementation for dot product computation. In this context, each thread calculates an individual vector. This approach is particularly efficient in the Vecchia framework, where vector sizes typically range from 30 to 120 elements, but the quantity of vectors extends to 500K or more. Consequently, allocating a single thread for each dot product operation is more effective than batched processing. 

    \item \textit{Data layout:} The KBLAS library  \cite{abdelfattah2016kblas} provides two data layouts: Point-to-Point (P2P) and strided. The strided layout is especially beneficial for the Vecchia algorithm for several reasons. First, it eliminates the need to merge previous data with new data points since all spatial or temporal data and observations are preloaded, avoiding the necessity for streaming access. Second, the strided layout ensures continuous data storage, enhancing GPU memory use efficiency. This is particularly important for managing the large-scale problems typical of the Vecchia algorithm. The data structure is shown in the Figure \ref{fig:data-structure}.
    
    \item \textit{Memory Allocation:} The main memory allocation challenge in our algorithm arises from the large number of dense, small matrices. The space complexity for this part of the algorithm is $\mathcal{O}(nm^2)$, where $m$ is the size of each small matrix, and $n$ is the total number of matrices. Besides, memory allocation for small vectors has space complexity of $\mathcal{O}(nm)$, which is particularly less impactful in the overall memory usage compared to the contiguous small matrices. These contiguous matrices and vectors are stored in the global memory of GPU.
\end{itemize}


\subsection{Memory Footprint and Arithmetic Complexity of Batched Vecchia}
\begin{figure*}[!h]
\centering
    \subfloat[(a) Memory complexity.]{\includegraphics[width=0.33\textwidth]{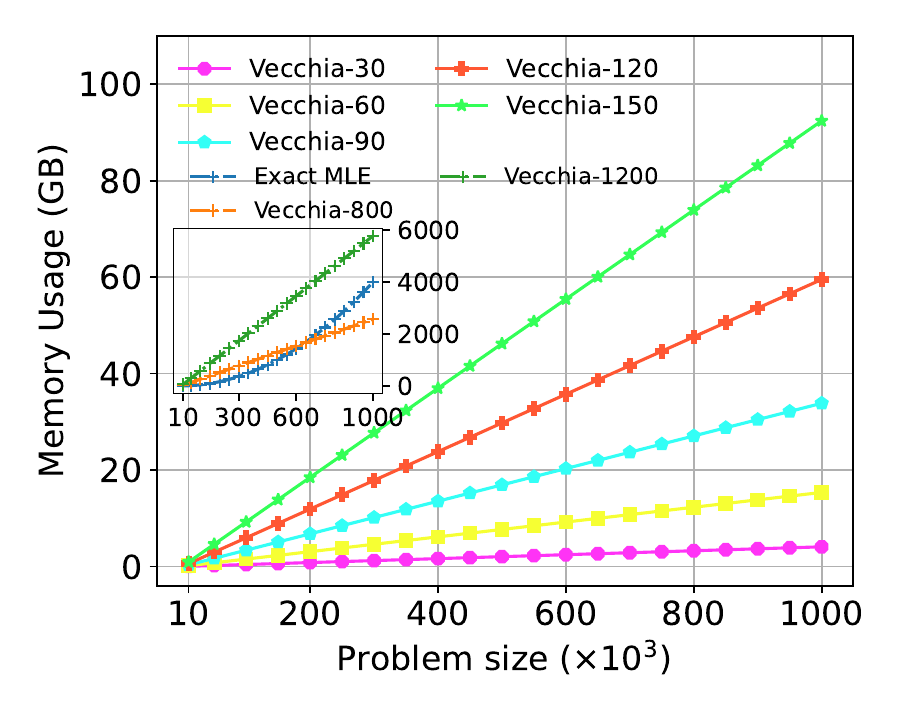}} 
    \subfloat[(b) Computational complexity.]{\includegraphics[width=0.33\textwidth]{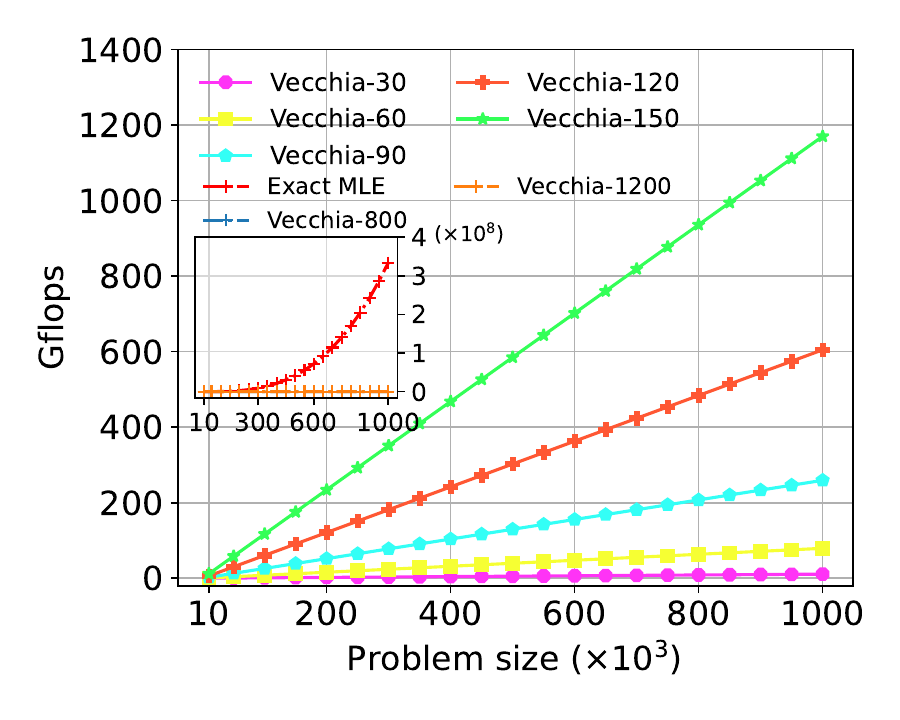} } 
    \caption{Comparison of Arithmetic complexity: Vecchia algorithm versus Exact MLE.}
    \label{fig:comp}
\end{figure*}

In this subsection, we analyze the memory footprint and the arithmetic complexity of the Batched Vecchia implementation proposed in {\it ExaGeoStat}, comparing it to the exact MLE implementation. The memory footprint of the exact MLE algorithm is $\sim{n^2/2}$ for the symmetric covariance matrix $\boldsymbol{\Sigma}$ and $\sim{n}$ for the measurements vector $\boldsymbol{y}$. For the Vecchia algorithm, each location requires a covariance matrix  $\boldsymbol{\Sigma}_i$, a vector $\boldsymbol{v}_i$, and a measurement vector $\boldsymbol{y}$, with memory complexities of $\sim{n (m^2/2)}$, $\sim{nm}$, and $\sim{nm}$, respectively. Figure ~\ref{fig:comp} (a) illustrates memory footprint in gigabytes (GB) for various problem sizes when employing the Vecchia and exact MLE algorithms. The figure emphatically highlights the benefits of utilizing the Vecchia approximation, which exhibits significantly lower memory requirements. The values 30, 60, 90, 120, and 150 represent the number of nearest neighbors considered for each location, which is represented by $m$.

The Vecchia algorithm significantly reduces the computational complexity compared to the exact MLE. For exact MLE, the complexity primarily stems from the Cholesky factorization of the covariance matrix, $\sim{n^3/3}$, and the triangular solve, $\sim{n^2}$. In contrast, with the Vecchia approximation, the complexity for the Cholesky factorization operations is $\sim{n(m^3/3)}$ (as shown in line 14 in Algorithm \ref{alg:batchedvecchia}), and there are triangular solves with vector $\boldsymbol{v}_i$ (line 15) and triangular solves with vector $\boldsymbol{y}_i$ (line 16). Figure ~\ref{fig:comp} (b) demonstrates the floating-point operations (flops) in Gflops for various problem sizes when using both the Vecchia and exact MLE algorithms. The figure shows a significant reduction in the number of required flops for the Vecchia algorithm, underlining its efficiency, particularly if it maintains accuracy comparable to the exact MLE solution.

In these two aforementioned subfigures, we also show the arithmetic complexity of large conditioning sets $m$ for the batched Vecchia that exceeds the memory requirement of the exact MLE. Figure~\ref{fig:comp} (a) shows that when $m=1200$, the memory requirement of the batched Vecchia algorithm exceeds the exact MLE algorithm with different problem sizes. However,  Figure~\ref{fig:comp} (b) shows that Vecchia still requires fewer flops than exact MLE at the same conditioning set size.

\section{Results and Discussions}
\label{sec:exp}

In this section, we conduct a series of experiments to evaluate the accuracy and performance of the batched Vecchia algorithm with several goals: (1) Numerically assess the accuracy of the batched Vecchia algorithm by comparing it to the exact MLE using KL divergence. (2) Evaluate the accuracy of the batched Vecchia algorithm using real datasets, focusing on modeling and prediction accuracy. (3) Examine the performance of the implemented algorithm across various NVIDIA GPUs, specifically the GV100, A100, and H100. (4) Investigate the largest problem size manageable with different conditioning sets in the Vecchia algorithm on various GPUs. (5) Discuss optimal parameters for the batched Vecchia approximation to achieve the best performance while maintaining the necessary accuracy.

\subsection{Experimental Testbed}
We conduct accuracy and performance assessment experiments using a range of GPUs, including a single NVIDIA GV100 with 32 GB of memory, a single NVIDIA A100 with 80 GB of memory, and an H100 with 80 GB of memory.

Our computational harness is built using gcc version 10.2.0 (12.2.0) and CUDA version 11.4 (11.8). It was linked with the KBLAS library, Intel MKL 2022.2.1, MAGMA 2.6.0, and NLopt v2.7.1 optimization libraries. All computations were performed in double-precision arithmetic, and we conducted each experimental run five times to verify repeatability. To assess accuracy and perform qualitative analysis, we utilized numerical calculations and examined two real datasets: the soil moisture dataset from the Mississippi River Basin region and the wind speed dataset from the Middle East region.

\subsection{Numerical Study}

In this subsection, we utilize exact log-likelihood calculated by {\em ExaGeoStat}~\cite{abdulah2018parallel}, focusing on Gaussian random fields with problem sizes of 180K and 260K in two-dimensional (2D) spatial locations. We use the Matérn kernel as described in (\ref{eq:maternkernel}) and vary the smoothness parameter $\nu$ at values of 0.5, 1.5, and 2.5, corresponding to low, medium, and high smoothness levels. Additionally, for each level of smoothness, we adjust the effective range to 0.1, 0.3, and 0.8 to account for low, medium, and high dependence values, respectively, which impacts data correlation. Detailed configurations can be found in Table \ref{tab:accuracy-setting} as part of our study on Gaussian random fields.

\begin{table}[h]
    \caption{The cross combinations of low/medium/high smoothness and low/medium/high effective range. Each entry in the table represents $\beta$, and the $0.1, 0.3, 0.8$ are the statistically effective range, i.e., the distance over which spatial dependencies are significant in the statistical model~\cite{huang2021competition}.}
    \centering
    \label{tab:accuracy-setting}
        \begin{tabular}{llll}
        \toprule
                              & $\nu=0.5$                        & $\nu=1.5$                       & $\nu=2.5$                        \\
                              \midrule
        effective range=$0.1$ & $0.026270$  & $0.017512$ & $0.014290$  \\
        effective range=$0.3$ & $0.078809$ & $0.052537$ & $0.014290$  \\
        effective range=$0.8$ & $0.210158$ & $0.140098$ & $0.114318$ \\\bottomrule
        \end{tabular}
\end{table}

\begin{figure}[t]
    \subfloat[$\beta=0.02627, \nu=0.5$]{\includegraphics[width=0.165\textwidth]{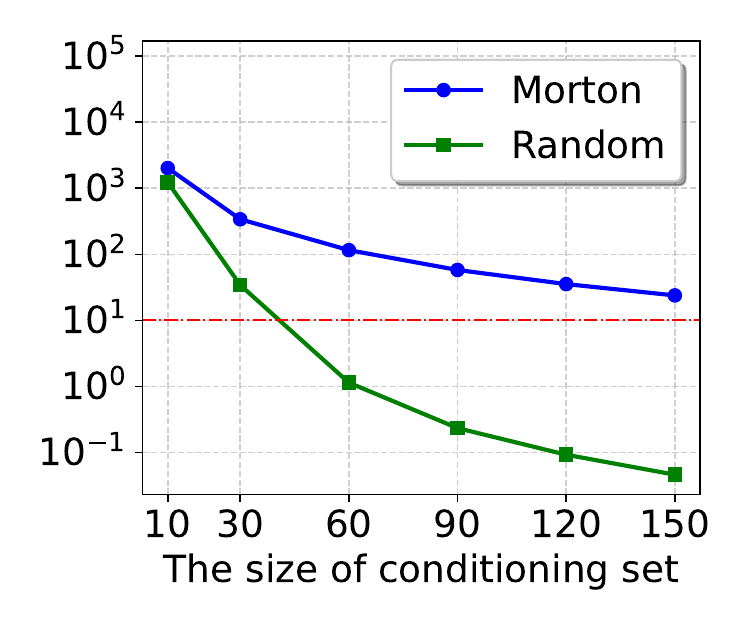}}    
    \subfloat[$\beta=0.017512, \nu=1.5$]{\includegraphics[width=0.165\textwidth]{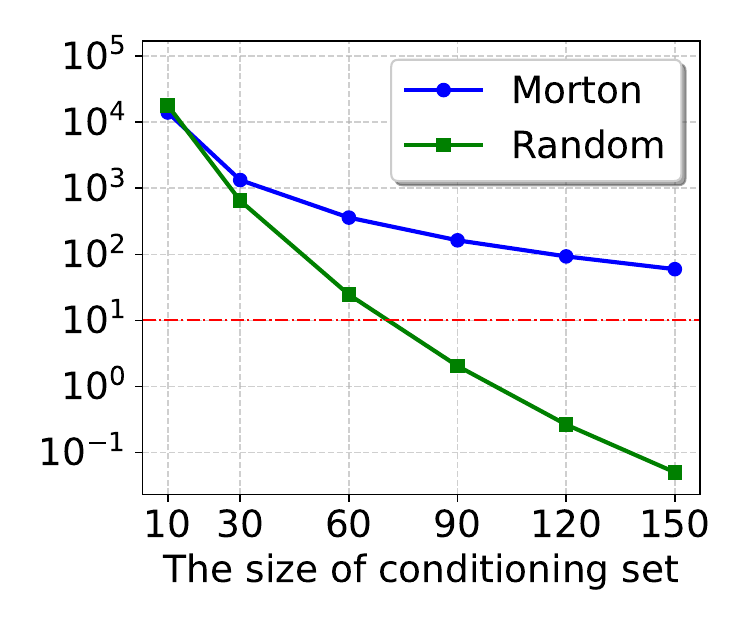}}    
    \subfloat[$\beta=0.01429, \nu=2.5$]{\includegraphics[width=0.165\textwidth]{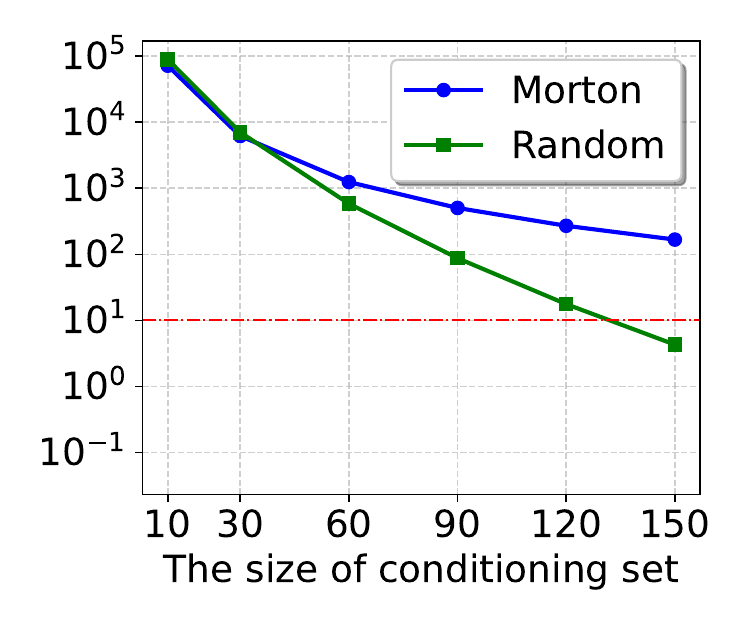}} \\    
    \subfloat[$\beta=0.078809, \nu=0.5$]{\includegraphics[width=0.165\textwidth]{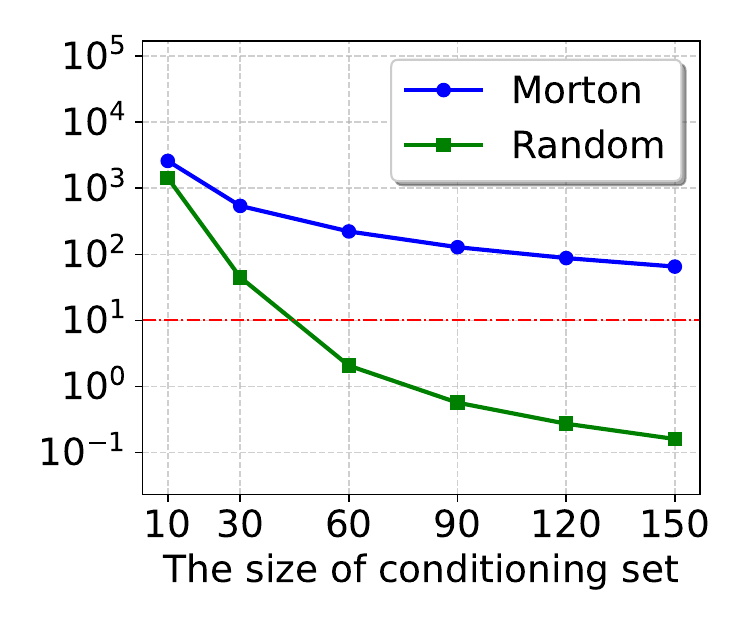}}    
    \subfloat[$\beta=0.052537, \nu=1.5$]{\includegraphics[width=0.165\textwidth]{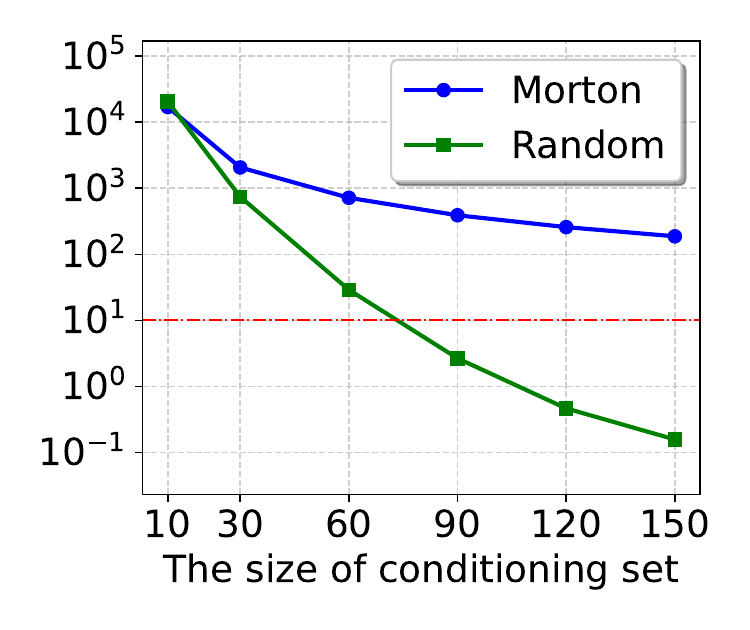}}    
    \subfloat[$\beta=0.042869, \nu=2.5$]{\includegraphics[width=0.165\textwidth]{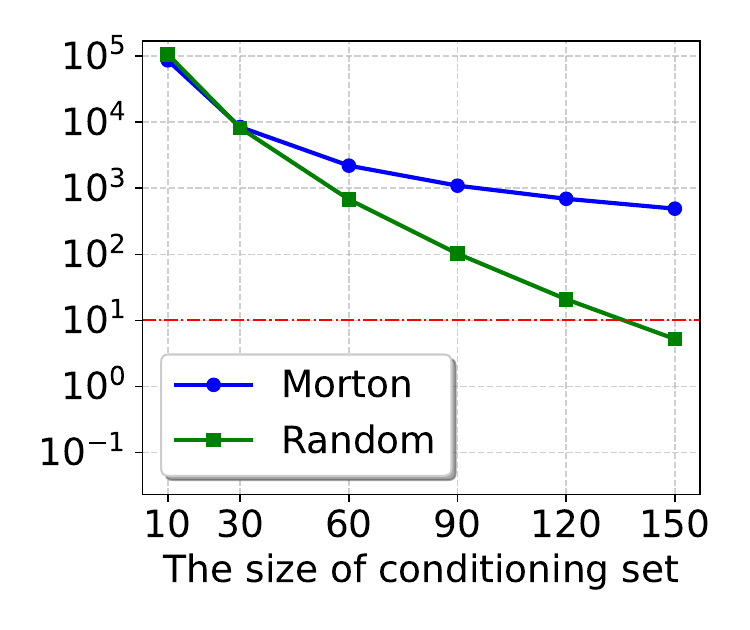}} \\    
    \subfloat[$\beta=0.210158, \nu=0.5$]{\includegraphics[width=0.165\textwidth]{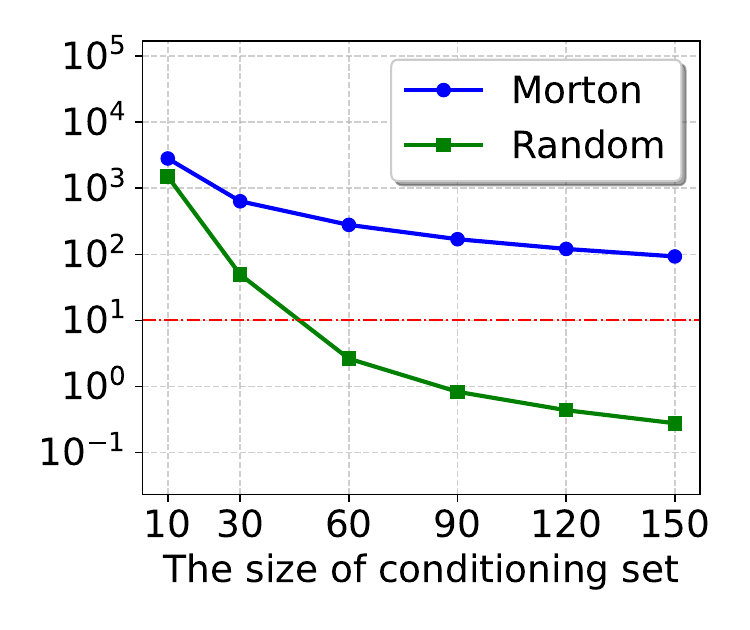}}    
    \subfloat[$\beta=0.140098, \nu=1.5$]{\includegraphics[width=0.165\textwidth]{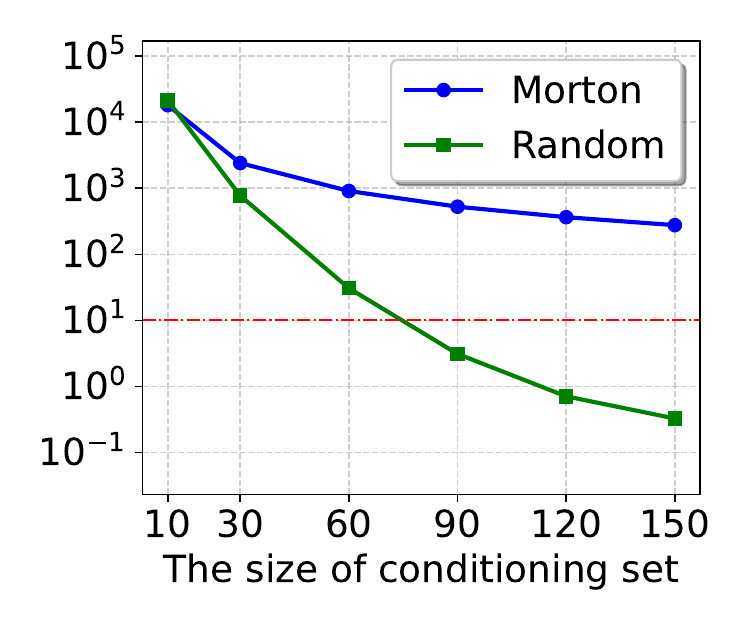}}    
    \subfloat[$\beta=0.114318, \nu=2.5$]{\includegraphics[width=0.165\textwidth]{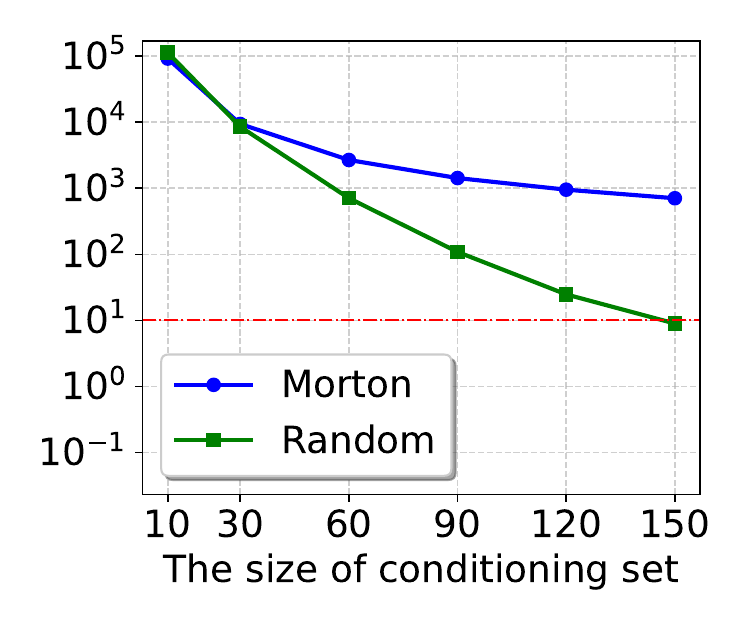}} \\   
    \caption{KL divergence under 260k locations with log10 scale (y-axis is KL divergence). The red dash line is the recommended threshold for choosing the conditioning size.}
    \label{fig:260kaccuracy}
\end{figure}

The calculation of the KL divergence is performed using (\ref{eq:kl-gaussian-vecchia}), and the outcomes are visualized in Figure \ref{fig:180kaccuracy} and Figure \ref{fig:260kaccuracy}. Across all subfigures, the x-axis corresponds to the size of the conditioning set, while the y-axis represents the KL divergence value, with $0$ indicating a perfect match with the exact 
MLE operation. Examination of these figures yields several significant observations:

\begin{figure}[t]
    \subfloat[$\beta=0.02627, \nu=0.5$]{\includegraphics[width=0.165\textwidth]{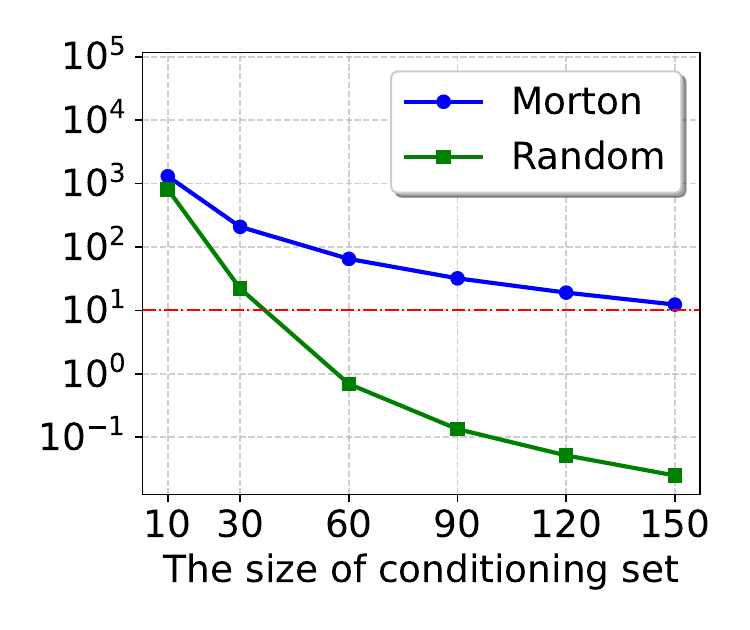}}    
    \subfloat[$\beta=0.017512, \nu=1.5$]{\includegraphics[width=0.165\textwidth]{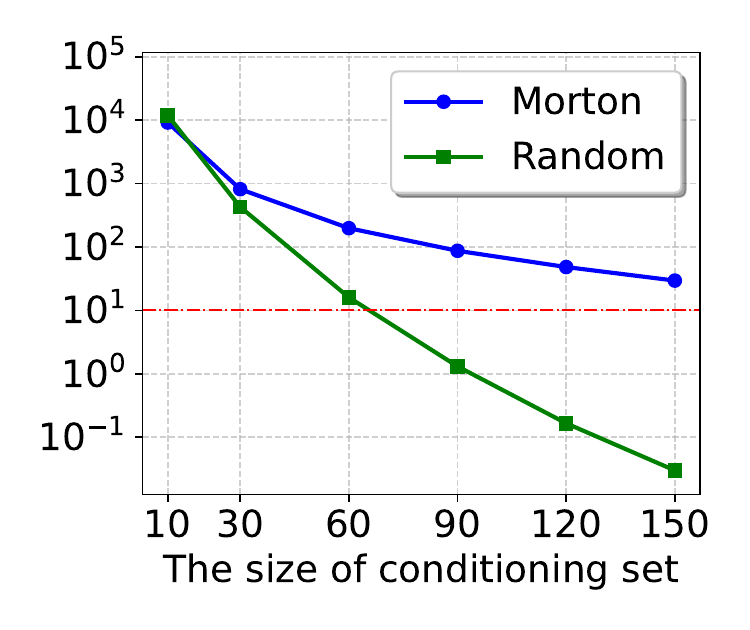}}    
    \subfloat[$\beta=0.01429, \nu=2.5$]{\includegraphics[width=0.165\textwidth]{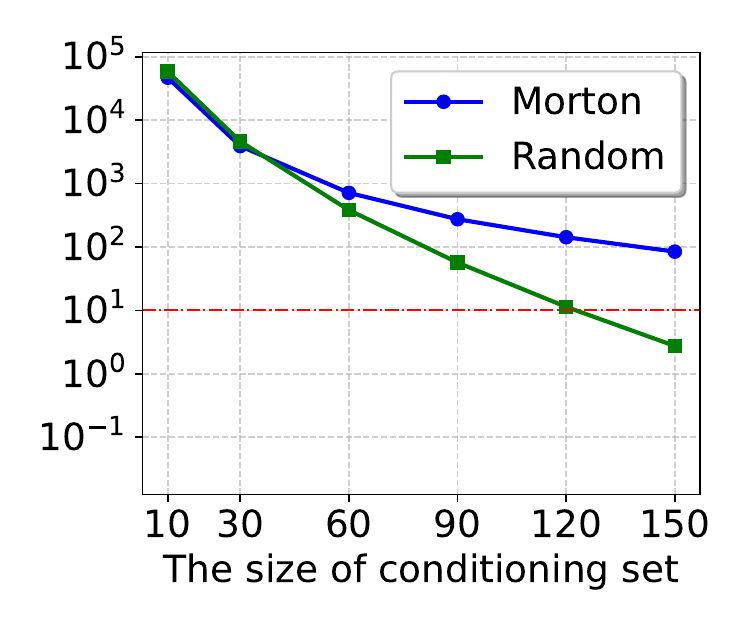}} \\    
    \subfloat[$\beta=0.078809, \nu=0.5$]{\includegraphics[width=0.165\textwidth]{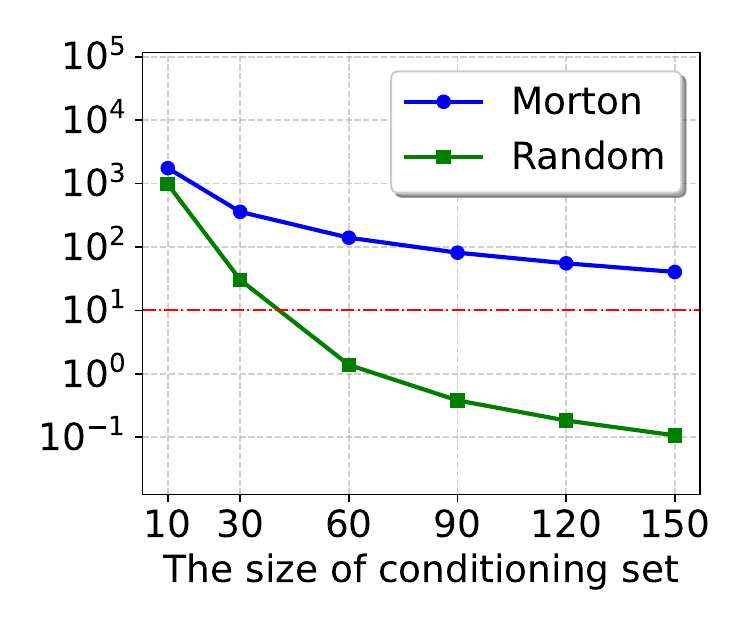}}    
    \subfloat[$\beta=0.052537, \nu=1.5$]{\includegraphics[width=0.165\textwidth]{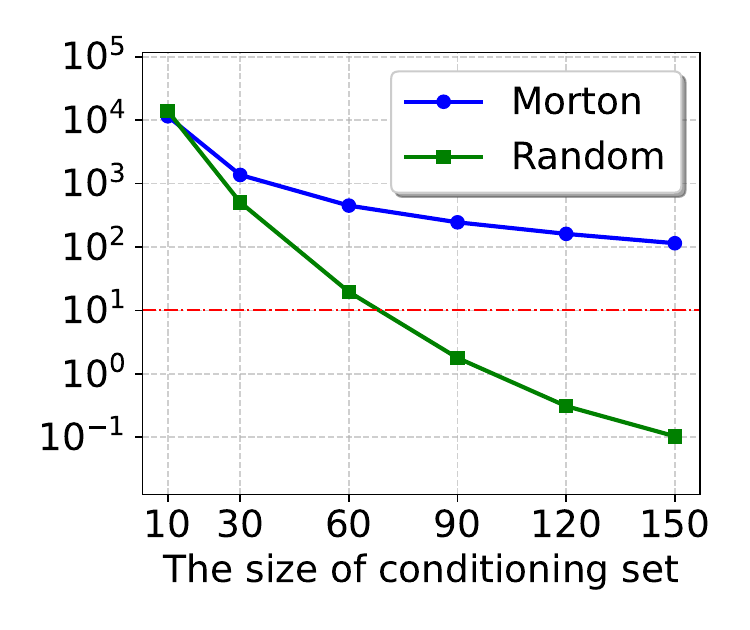}}    
    \subfloat[$\beta=0.042869, \nu=2.5$]{\includegraphics[width=0.165\textwidth]{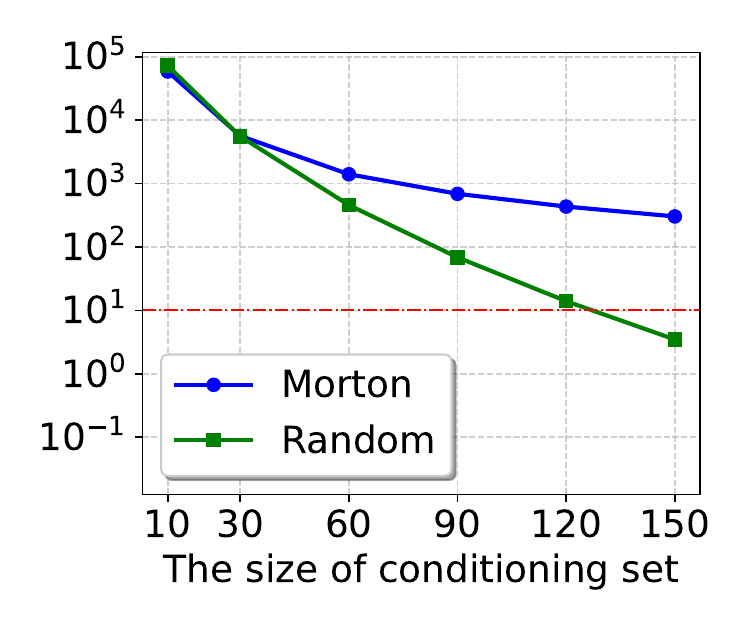}} \\    
    \subfloat[$\beta=0.210158, \nu=0.5$]{\includegraphics[width=0.165\textwidth]{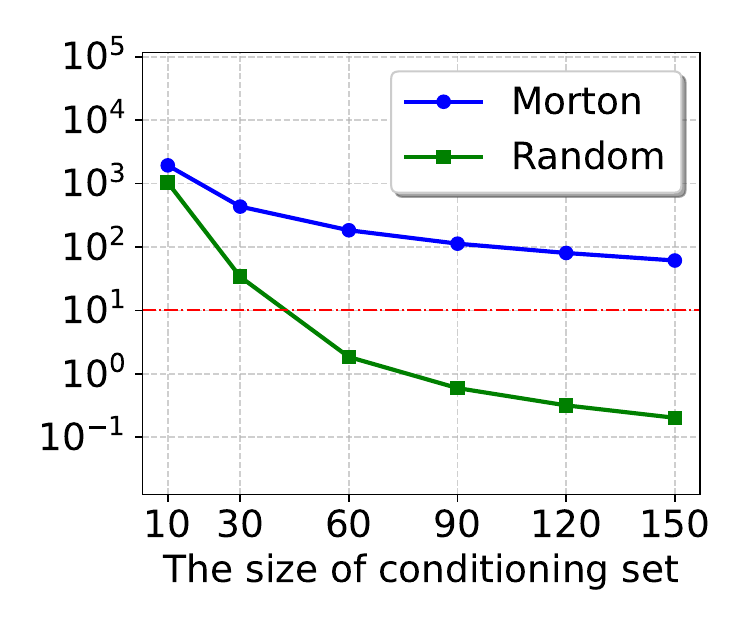}}    
    \subfloat[$\beta=0.140098, \nu=1.5$]{\includegraphics[width=0.165\textwidth]{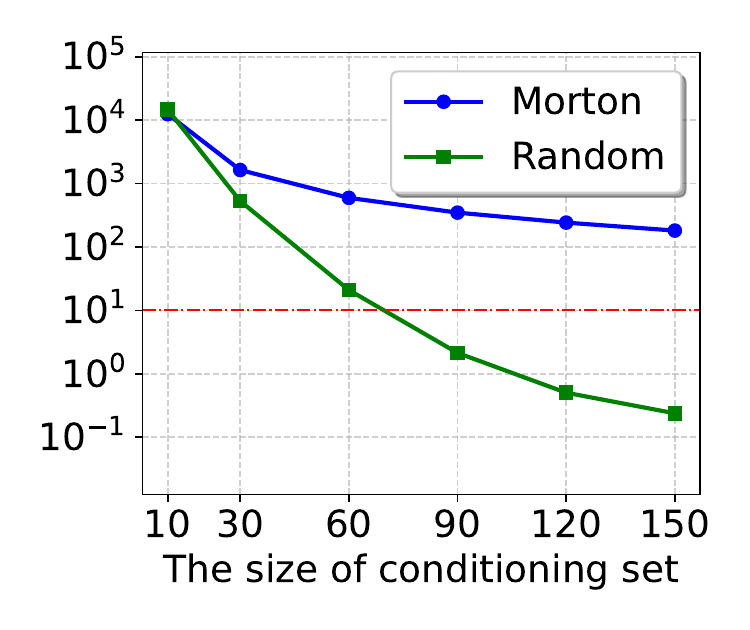}}    
    \subfloat[$\beta=0.114318, \nu=2.5$]{\includegraphics[width=0.165\textwidth]{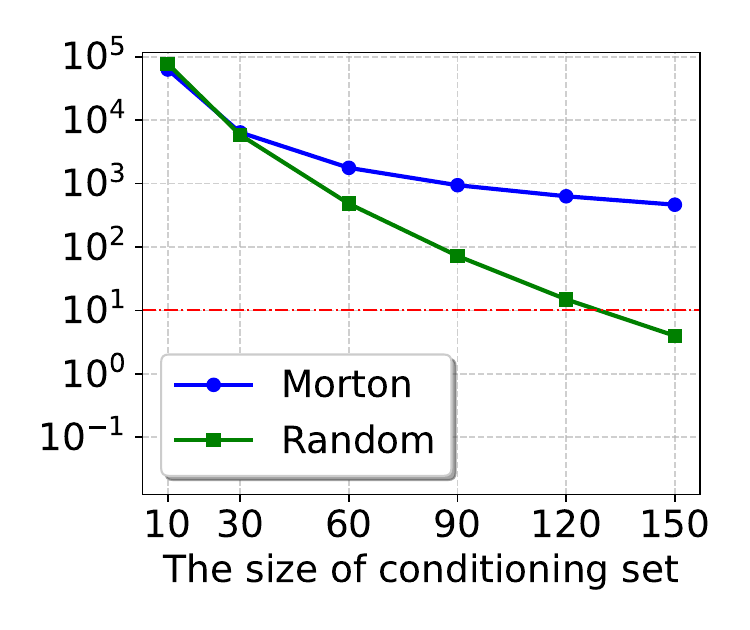}} \\   
    \caption{KL divergence under 180K locations with log10 scale (y-axis is KL divergence). The red dashed line is the recommended threshold for choosing the conditioning size.}
    \label{fig:180kaccuracy}
\end{figure}

\begin{itemize}
    \item The significance of spatial ordering in log-likelihood approximation cannot be overstated. When it comes to accuracy, it becomes evident that random ordering outperforms Morton's ordering at large-scale problems. This observation highlights the role of selecting the right ordering strategy for achieving effective approximations.

    \item Impact of range and smoothness on approximation difficulty. The complexity of the approximation escalates with increases in range or smoothness parameters. A notable enlarged KL divergence under elevated range or smoothness conditions evidences this. Consequently, the additional conditioning points become imperative in the high-range or high-smoothness scenarios.
    
    \item Approximation challenge for large-scale problem size. The difficulty of approximation is directly proportional to the size of the problem. The numerical study reveals an increase in KL divergence when the problem size escalates from 180K to 260K. This observation suggests that larger problem sizes necessitate the use of more conditioning points to maintain approximation accuracy.
\end{itemize}

\subsection{Accuracy Assessment of Batched Vecchia with Real Data}
\subsubsection{Soil Moisture}

\begin{figure}[h]
    \centering
    \subfloat[Residual of Observations (soil)]{\label{fig:soil}\includegraphics[width=0.4\textwidth]{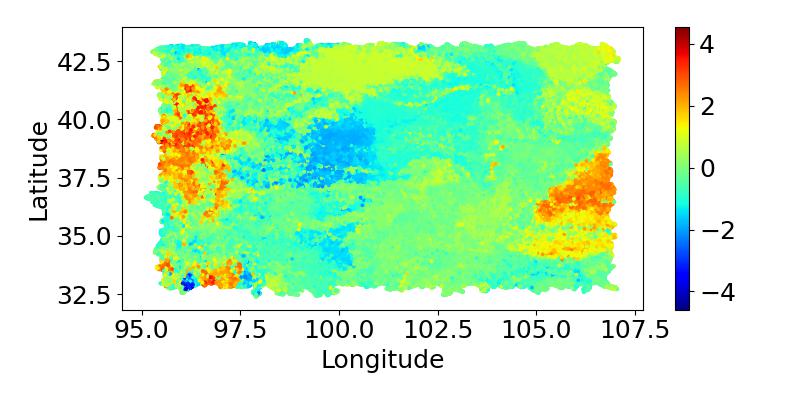}}\\
    \subfloat[Residual of Observations (wind \textcolor{black}{speed})]
    {\label{fig:wind}\includegraphics[width=0.4\textwidth]{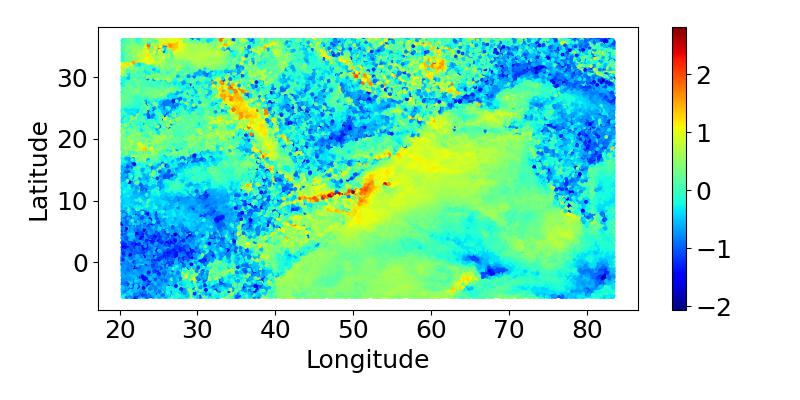}}
    \caption{Real datasets residuals: soil moisture and wind speed, \textcolor{black}{where the observations of soil moisture and the square root of wind speed served as the response variables, respectively.}}
    \label{fig:dataset}
\end{figure}

This study examines high-resolution daily soil moisture data obtained from the Mississippi River basin in the United States on January 1, 2004, as reported in \cite{chaney2016hydroblocks}. The dataset, which was previously utilized in \cite{huang2018hierarchical} and \cite{abdulah2018exageostat}, involves Gaussian field modeling and contains 2 million irregularly distributed locations. To manage computational costs, we randomly selected 250K locations as the training dataset and 25K as the testing dataset. This choice allows us to compare the estimated parameter vector and predictions obtained via the Vecchia approximation with those from exact modeling, where using all 2 million locations would be computationally burdensome. The data has a spatial resolution of 0.0083 degrees, with each one-degree difference approximately corresponding to a distance of 87.5 km. Consequently, we utilize the Great Circle Distance (GCD) metric to calculate the distances between any pair of locations based on their original longitude and latitude values, as described in \cite{abdulah2018exageostat}:
$$
\left(\frac{d}{r}\right)=\operatorname{hav}\left(\varphi_2-\varphi_1\right)+\cos \left(\varphi_1\right) \cos \left(\varphi_2\right) \operatorname{hav}\left(\lambda_2-\right.
\left.\lambda_1\right).
$$
Here, the haversine function denoted as $\operatorname{hav}(\cdot)$, is defined as $\sin^2\left(\frac{\cdot}{2}\right)=\frac{1-\cos(\cdot)}{2}$, where $d$ represents the distance between two locations, $r$ is the radius of the sphere, $\varphi_1$ and $\varphi_2$ are the latitudes in radians of locations 1 and 2, respectively, and $\lambda_1$ and $\lambda_2$ are their respective longitudes.

The mean value of the raw dataset is removed by a linear regression model where the response variable is observation, and explanatory variables are longitude and latitude. The residuals, as shown in Figure \ref{fig:dataset}, are fitted using a zero-mean Gaussian process model, which incorporates a power exponential covariance function,
\begin{equation*}
    C(d) = \sigma^2 \exp\{-d^\alpha/\beta\},
\end{equation*}
where $d$ is the distance between two locations and the parameter vector $\boldsymbol\theta = (\sigma^2, \beta, \alpha)^\top$  represent the variance, range and smoothness, respectively. For the Vecchia approximated Gaussian process, six different conditioning sizes (10, 30, 60, 90, 120, 150) with random ordering and 250K subsampling problem size are considered. 
We use {\it ExaGeostat} \cite{abdulah2018exageostat} to estimate the parameter vectors for the exact Gaussian process. In the end, the estimated parameter vectors are plugged into the kriging, and then we obtain the Mean Square Error (MSE) for the prediction task.

\subsubsection{Wind Speed}
The WRF-ARW (Weather Research and Forecasting - Advanced Research WRF) model generated a regional climate dataset specific to the Arabian Peninsula in the Middle East, as documented in \cite{powers2008description}. The model is configured with a horizontal grid spacing of 5 km, encompassing 51 vertical levels, and the highest level of the model is established at 10 hPa. Geographically, the model's domain covers the area from 20°E to 83°E in longitude and from 5°S to 36°N in latitude. This dataset spans 37 years, with daily data provided. Each data file contains a complete day's (24 hours) record of hourly wind speed measurements across 17 distinct atmospheric layers. For this study, we focus on the dataset from September 1, 2017, starting at 00:00 AM. Our interest is in wind speed measurements at a height of 10 meters above the ground, corresponding to the lowest layer, referred to as layer 0. The method of calculating distances in the wind speed dataset is consistent with that used in the soil moisture dataset. The residuals of wind speed are visualized in Figure \ref{fig:dataset}, where we utilized the square root of wind speed as the response variable and longitude and latitude as explanatory variables, taking into account the skewed distribution of wind speed—additionally, the same experimental settings as the soil moisture dataset were applied. Herein, the initial dataset comprises 1M locations, from which we randomly extracted 250K locations for training and 25K for testing.

\subsubsection{Result Analysis}
\begin{figure*}[h]
    \centering

    \subfloat[Soil $\hat \sigma^2$]{
    \label{fig:soil-sigma}
    \includegraphics[width=0.30\textwidth]{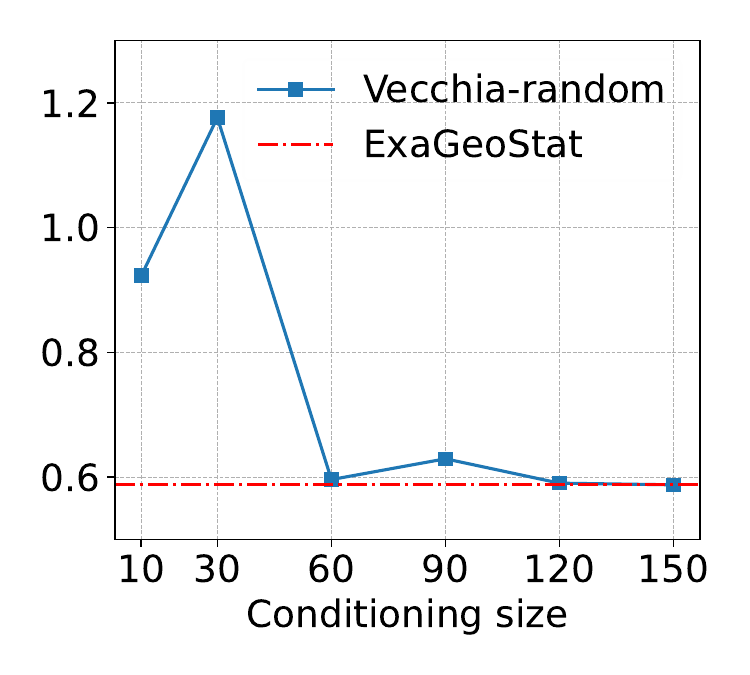}
    }
    \subfloat[Soil $\hat \beta$]{
    \label{fig:soil-beta}
    \includegraphics[width=0.30\textwidth]{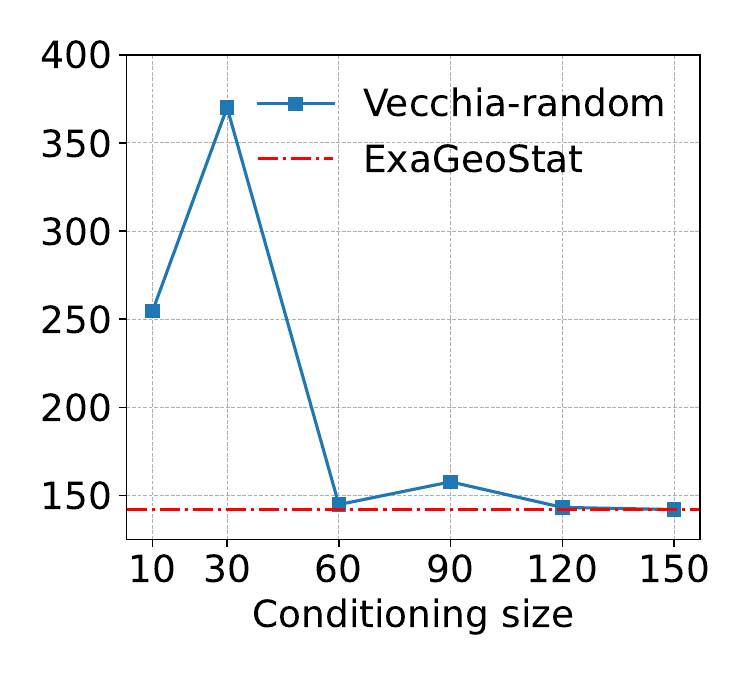}
    }
    \subfloat[Soil $\hat \alpha$]{
    \label{fig:soil-alpha}
    \includegraphics[width=0.30\textwidth]{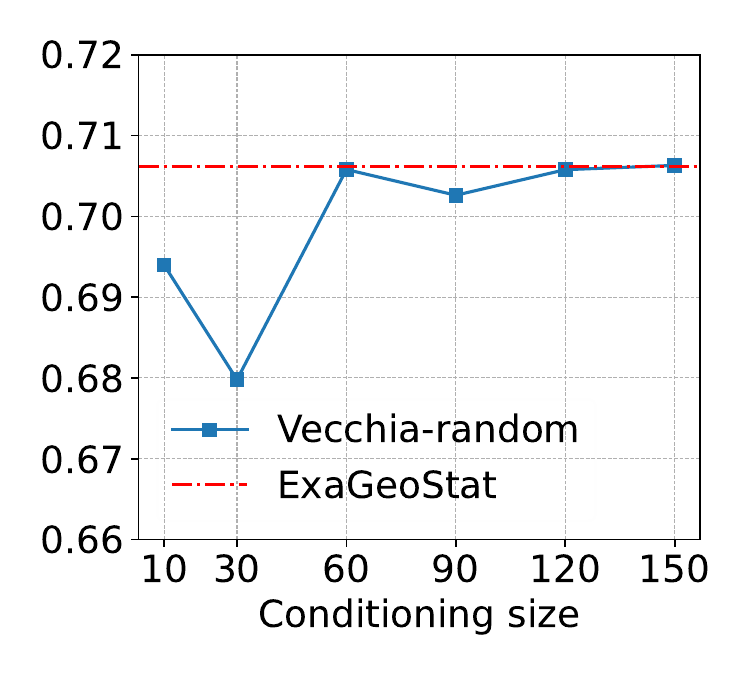}
    }
    \\
    \subfloat[Wind $\hat \sigma^2$]{
    \label{fig:wind-sigma}
    \includegraphics[width=0.30\textwidth]{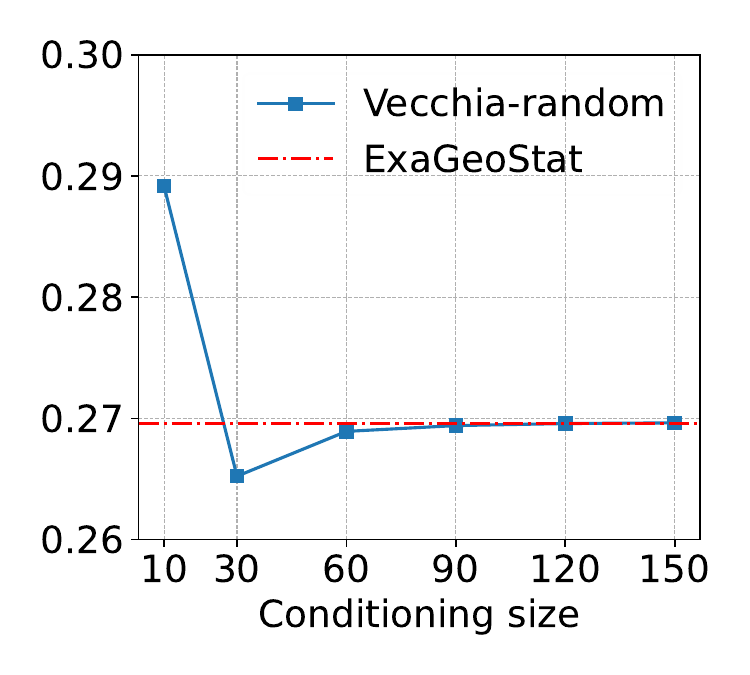}
    }
    \subfloat[Wind $\hat \beta$]{
    \label{fig:wind-beta}
    \includegraphics[width=0.30\textwidth]{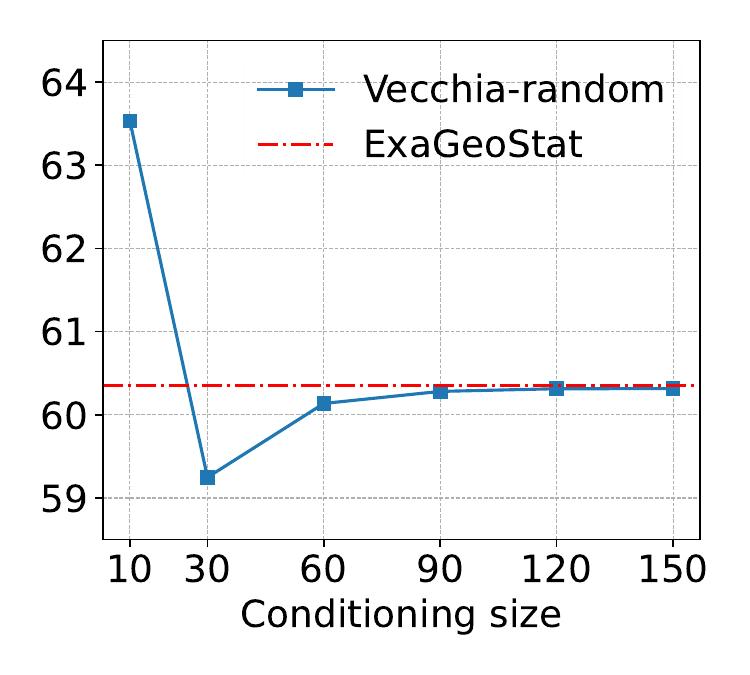}
    }
    \subfloat[Wind $\hat \alpha$]{
    \label{fig:wind-alpha}
    \includegraphics[width=0.30\textwidth]{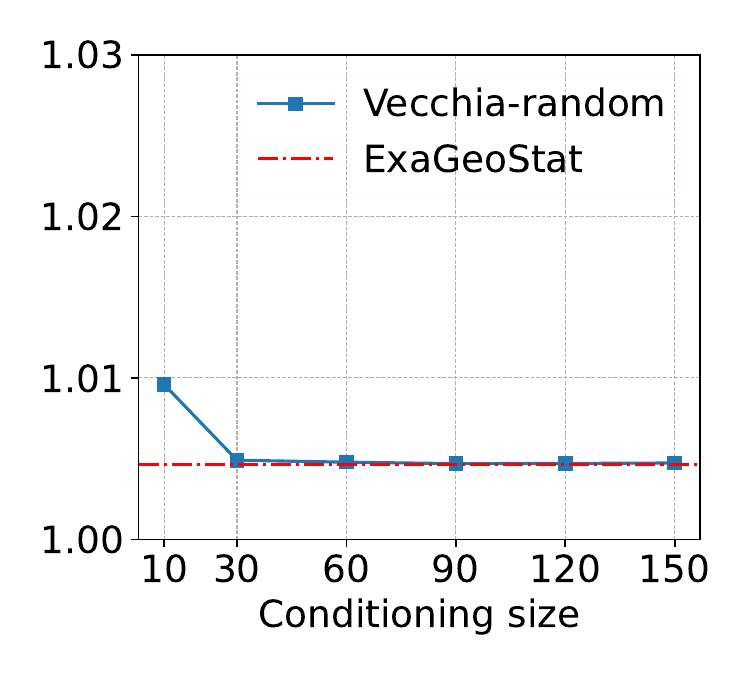}
    }
    
    \caption{The estimated parameter vectors using Vecchia approximation with different conditioning sizes compared to  {\it ExaGeoStat} (exact MLE). The first row is the parameter vector for soil moisture, and the second for wind speed.}
    \label{fig:realapp-results}
\end{figure*}

Figure \ref{fig:realapp-results} shows the estimated parameter vectors for both datasets, while Table \ref{tab:pre-mse} highlights the MSE associated with the prediction. During the estimation process, it was observed that the parameter vector $\boldsymbol\theta$, as estimated through the Vecchia approximation, closely aligns with that obtained via {\it ExaGeoStat} (exact MLE), particularly as the number of conditioning neighbors increases. Figure \ref{fig:realapp-results} illustrates that, for both datasets, a conditioning size of 60 is optimal for achieving an estimation close to the exact MLE. Table \ref{tab:pre-mse} further demonstrates that the Vecchia approximation achieves a prediction error remarkably close to the actual values when utilized for predicting missing data. 

\begin{table}[]
\centering
\caption{MSE of {\it ExaGeoStat} and Vecchia on soil and wind dataset ($\times 10^{-2}$).}
\label{tab:pre-mse}
\begin{tabular}{ccccccccc}
\toprule
\multirow{2}{*}{} & \multicolumn{2}{c}{\multirow{2}{*}{\begin{tabular}[c]{@{}c@{}}{\it ExaGeoStat} \\ (Exact MLE)\end{tabular}}} & \multicolumn{4}{c}{Vecchia} \\
                  & \multicolumn{2}{c}{}    & 60  & 90  & 120 & 150 \\ \midrule
Soil      & 7.832727 &    & 7.850310 & 7.849056 & 7.850950 &7.850904   \\
Wind         & 2.842985 &    &  2.842939&  2.842972& 2.842967  &2.842954   \\ \bottomrule

\end{tabular}
\end{table}

\subsection{Performance Assessment}
\label{sec:perf}
\begin{figure*}[htbp]
\centering
\subfloat[NVIDIA GV100 GPU with 32GB.]{\includegraphics[width=0.30\textwidth]{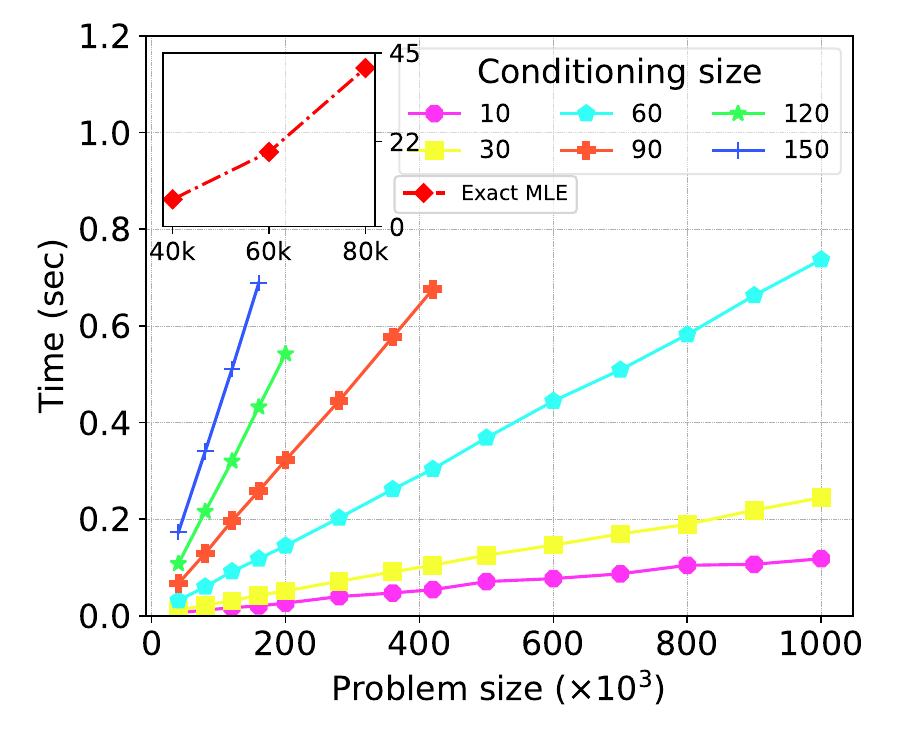}}
\hspace{4mm}
\subfloat[NVIDIA A100 GPU with 80GB.]{\includegraphics[width=0.30\textwidth]{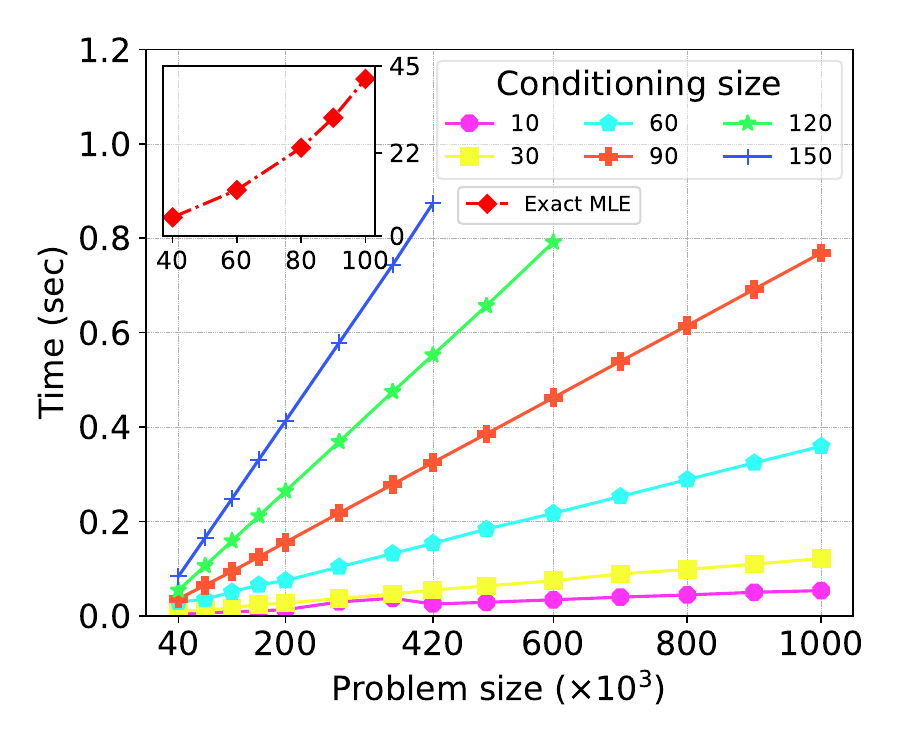}}
\hspace{4mm}
\subfloat[NVIDIA H100 GPU with 80GB.]{\includegraphics[width=0.30\textwidth]{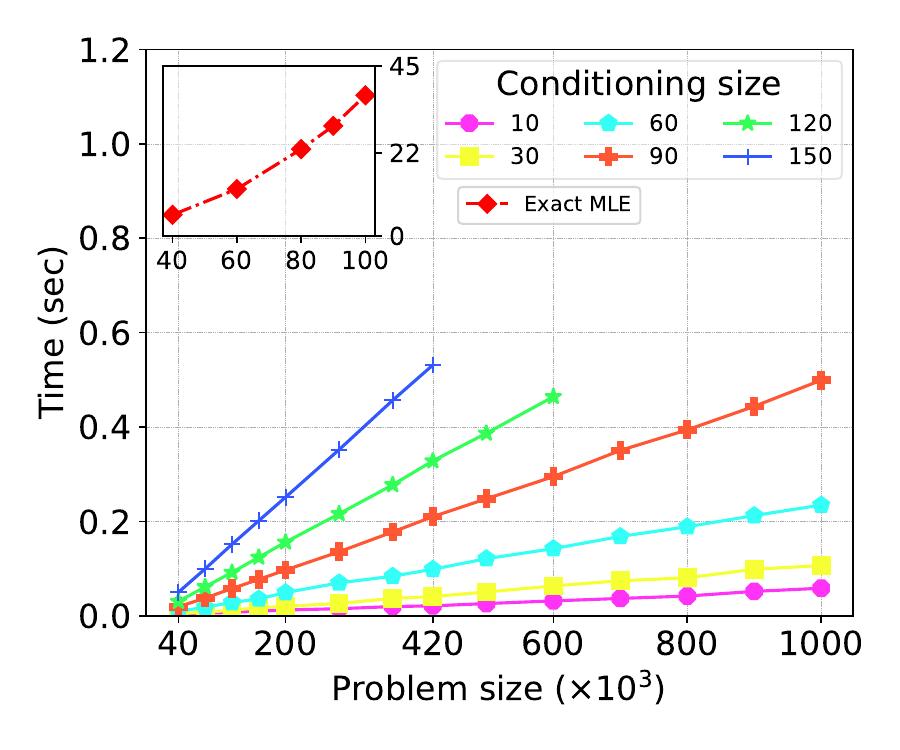}}\\
\subfloat[NVIDIA GV100 GPU with 32GB.]{\includegraphics[width=0.30\textwidth]{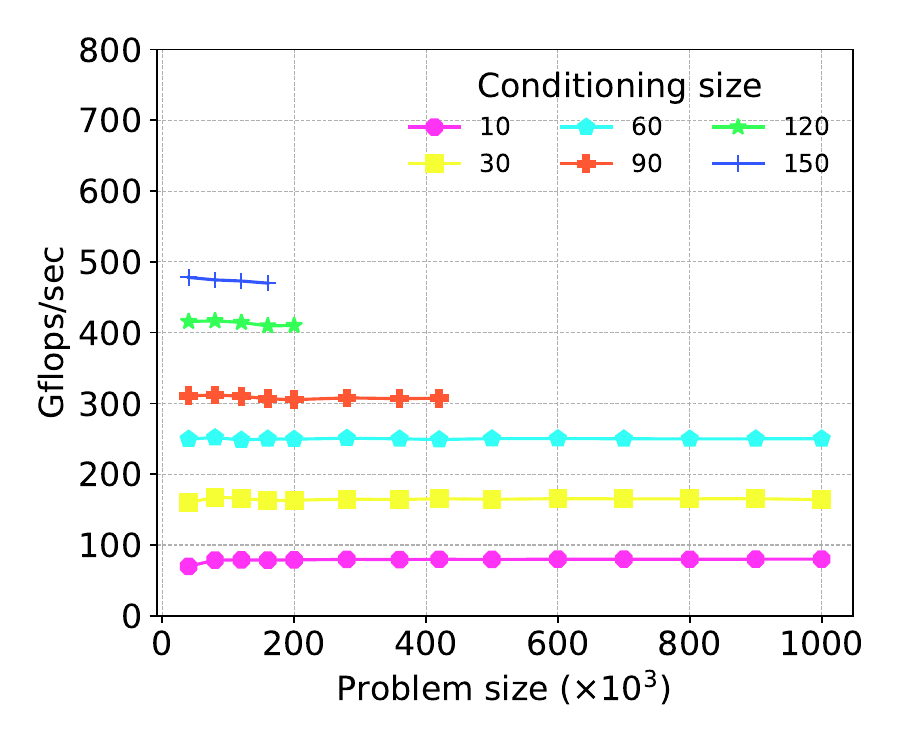}}
\hspace{4mm}
\subfloat[NVIDIA A100 GPU with 80GB.]
{\includegraphics[width=0.30\textwidth]{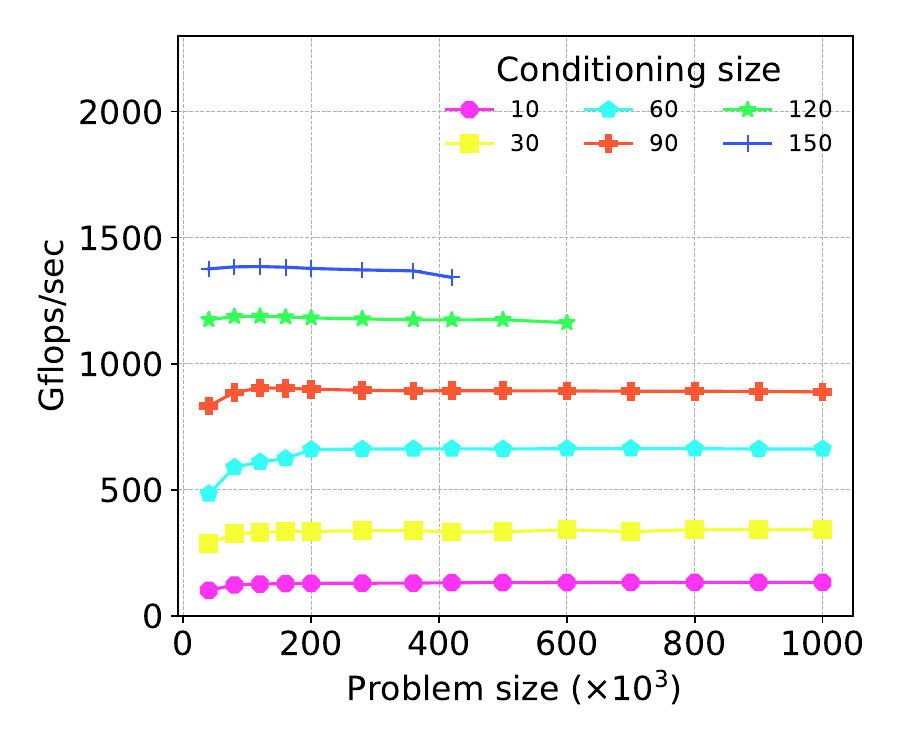}}
\hspace{4mm}
\subfloat[NVIDIA H100 GPU with 80GB.]{\includegraphics[width=0.30\textwidth]{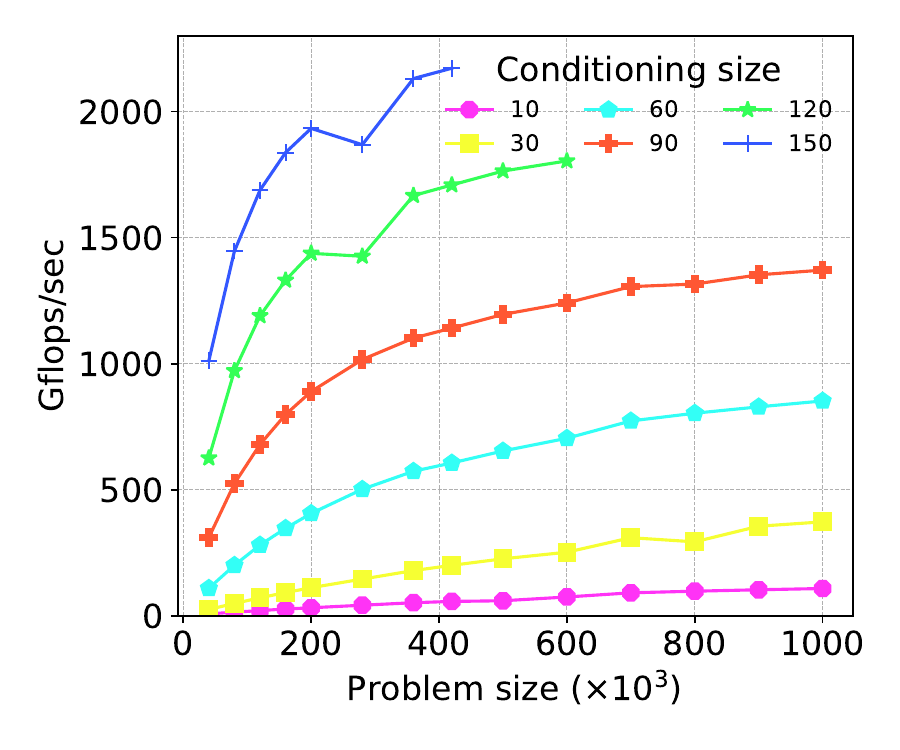}}
     
\caption{Evaluation of computational performance on various NVIDIA GPUs, i.e., GV100, A100, and H100: the first row indicates the execution time required for a single likelihood estimation, \textcolor{black}{where the small subfigures represent the performance of \it{ExaGeoStat}}. The second row shows the Gflops/sec achieved by our implementation for a single likelihood estimation.}
\label{fig:perf}
\end{figure*}

In this study, we evaluate the performance of batched Vecchia approximation across three distinct GPU architectures: GV100 (Quadro Volta) with 32 GB, A100(Ampere) with 80 GB, and H100 (Hopper) with 80 GB. The results presented in this subsection represent the average of five separate runs. Our objective is to comprehensively understand the efficacy of our software under varying computational conditions. To this end, we employ a range of conditioning sizes (batch sizes), specifically 
$(10, 30, 60, 90, 120, 150)$. The selection of smaller sizes is informed by the recommendation of approximately 30 neighbors for optimal Vecchia approximation, as suggested by \cite{guinness2018permutation}. Larger values are chosen in response to the demands of extensive problem sizes, which may necessitate increased conditioning sizes. Additionally, we adjust the problem size for each GPU to leverage their memory capacities fully. This approach allows us to explore the upper limits of computational efficiency within the constraints of each hardware configuration. The analysis includes assessing the batched Vecchia method against {\it ExaGeoStat}-GPU. This comparison focuses on two primary metrics: time and Gflops/sec. Time encompasses the total duration required for computing the log-likelihood, which includes the generation of the covariance matrix and associated log-likelihood operations (POTRF/TRSM/dot product). The Gflops/sec metric specifically evaluates the efficiency of log-likelihood-related operations. The results are depicted in Figure \ref{fig:perf}, from which several significant insights are discerned:

\begin{itemize}
\item \textit{Linear relationship between time and problem size:} Within the same hardware, we observe a linear escalation in time correlating with increases in problem size. This finding provides a useful framework for predicting computational time across varying problem scales.

\item \textit{Comparison with {\it ExaGeoStat}-GPU:} When evaluated on the metric of a single likelihood estimation time, the Vecchia method \textcolor{black}{with 60 neighbors} outperforms {\it ExaGeoStat}-GPU, exhibiting an approximately \textcolor{black}{700X, 833X, 1380X speedups compared to exact MLE, where the problem size is the largest matrix dimension that can fully fit into the GPU memory for exact MLE}. Moreover, considering the space complexity, the batched Vecchia approximation can handle larger problem sizes of up to 1 million in a single GPU.

\end{itemize}

\section{Conclusions}
\label{sec:conclusion}
Gaussian processes (GPs) are a powerful and flexible tool used in statistical modeling and machine learning for various tasks, including modeling, regression, and classification. However, GPs encounter a significant computational burden when dealing with high-dimensional data, prohibiting its use with massive amounts of data. Thus, many methods have been proposed to approximate the covariance matrix associated with the GPs. Among these methods is the Vecchia approximation algorithm, which allows a large-scale approximation of GPs. 

This work presents a parallel implementation of the Vecchia approximation technique that utilizes batched matrix computations on modern GPUs. Using batched linear algebra operations and the KBLAS library, the proposed implementation significantly reduces the time to solution compared to the state-of-the-art parallel implementation in the {\em ExaGeoStat} software. The speedup achieved on various GPU models (GV100, A100, H100) ranges from \textcolor{black}{ 700X to 1380X } compared to the exact solution. The implementation can also manage larger problem sizes, accommodating up to 1 million geospatial locations with 80GB A100 and H100 GPUs while maintaining accuracy. The study also assesses the accuracy performance of the Vecchia approximation algorithm on real geospatial datasets, specifically soil moisture data in the Mississippi Basin area and wind speed data in the Middle East. 
The code can be found at: \url{https://github.com/kaust-es/ParallelVecchiaGP}.

\section*{Acknowledgment}
This research received support from the King Abdullah University of Science and Technology (KAUST) in Saudi Arabia. Our gratitude extends to the team at the KAUST Supercomputing Laboratory (KSL) and the Extreme Computing Research Center (ECRC) for providing the computational resources that were essential for the experiments conducted in this study. \textcolor{black}{ We extend our gratitude to Jie Ren (ECRC/KAUST) and Mohsin Shaikh (KSL/KAUST) for their assistance throughout the project}.

\bibliographystyle{IEEEtran}
\bibliography{main}

\end{document}